\documentclass[aps,preprintnumbers,superscriptaddress,runinaddress,amsmath,amssymb,showpacs,floatfix,twocolumn]{revtex4-1}

\usepackage{graphicx}
\usepackage{dcolumn}
\usepackage{bm}
\usepackage{amsmath}
\usepackage{hyperref}
\usepackage{xcolor}
\usepackage{subfig}

\usepackage{caption}
\begin{document}

\preprint{ADP-13-06/T826}

\title{Structure and Flow of the Nucleon Eigenstates in Lattice QCD}

%=========================================

\author{M. Selim Mahbub} \author{Waseem Kamleh} \author{Derek
  B. Leinweber} \author{Peter J. Moran} \author{Anthony G. Williams}
\affiliation{Special Research Centre for the Subatomic Structure of
  Matter, School of Chemistry \& Physics, University of Adelaide,
  South Australia 5005, Australia.}

\collaboration{CSSM Lattice Collaboration}
\noaffiliation

\begin{abstract}
A determination of the excited energy eigenstates of the nucleon,
$s=\frac{1}{2}$, $I=\frac{1}{2}$, $N^{\pm}$, is presented in full QCD
using $2+1$ flavor PACS-CS gauge configurations. The
correlation-matrix method is used and is built using standard nucleon
interpolators employing smearings at the fermion sources and sinks. We
develop and demonstrate a new technique that allows the eigenvectors
obtained to be utilized to track the propagation of the intrinsic
nature of energy-states from one quark mass to the next.  This
approach is particularly useful for larger dimension correlation
matrices where more near-degenerate energy-states can appear in the
spectrum.
\end{abstract}


\maketitle

\section{ \label{sec1:intro}Introduction}

Resonances represent some of the rich dynamics of one of the
fundamental interactions of Nature, the strong interaction of quarks
and gluons.  Lattice QCD is the only ab initio first principles
approach to the fundamental quantum field theory governing the
properties of hadrons and ultimately we wish to test our theoretical
understanding of resonances against their experimentally determined
properties.

From lattice QCD, the ground-state hadron spectrum is now relatively
well understood~\cite{Durr:2008zz}. However, gaining knowledge of the
excited-state spectrum on the lattice presents additional challenges,
as the excited energy-states are extracted from the sub-leading
exponentials of the correlation functions. A determination of the
excited state energy spectrum, including
  multi-particle states, requires significant effort and some
progress is now being made. We can expect the interplay between
lattice QCD predictions and experimental measurement to be very
productive in the coming years.

In the case of nucleon resonances, the first positive parity excitation
of the nucleon, the $\rm{N}{\frac{1}{2}}^+(1440)\, {\rm{P}}_{11}$ or
Roper resonance, has been a subject of considerable interest since its
discovery in 1964 through a partial-wave analysis of pion-nucleon
scattering data~ \cite{Roper:1964zz}. This state has a surprisingly low
mass, which is well below the first negative parity excitation.
In constituent quark models with a harmonic oscillator potential this
${\rm{P}}_{11}$ state (with principal quantum number $\rm{N}=2$)
appears above the lowest-lying odd-parity ${\rm{S}}_{11}$ (1535)
state~\cite{Isgur:1977ef,Isgur:1978wd}, whereas in Nature the Roper
resonance is almost 100 MeV below the ${\rm{S}}_{11}$ state. This
presents a phenomenological challenge to our understanding of level
ordering. Similar difficulties occur with the $J^{P}={\frac{3}{2}}^{+}
\Delta^{\ast}(1600)$ and ${\frac{1}{2}}^{+} \Sigma^{\ast} (1690)$
resonances. Due to its surprisingly low mass, the ${\rm{P}}_{11}$
state has lead to enormous curiosity and much speculation about its
nature. For example, the Roper resonance has been described as a
hybrid baryon with explicitly excited gluon field
configurations~\cite{Li:1991yba,Carlson:1991tg}, or as a breathing
mode of the ground state~\cite{Guichon:1985ny} or as a five quark
(meson-baryon) state~\cite{Krehl:1999km}.
Significant resources have been devoted in the past from the lattice
QCD perspective to find the elusive low-lying Roper state, in both
quenched~\cite{Allton:1993wc,Lee:1998cx,Gockeler:2001db,Sasaki:2001nf,Melnitchouk:2002eg,
  Edwards:2003cd,Lee:2002gn,Brommel:2003jm,Mathur:2003zf,Sasaki:2003xc,Burch:2004he,Basak:2007kj,Mahbub:2009nr,Mahbub:2009aa,Fleming:2009wb,Mahbub:2010jz,Mahbub:2010me}
and in
full~\cite{Bulava:2009jb,Engel:2010my,Edwards:2011jj,Mahbub:2010rm,Lin:2011da}
QCD.

The `Variational method'~\cite{Michael:1985ne,Luscher:1990ck} is the
state-of-the-art approach for determining the excited state hadron
spectrum. It is based on the creation of a matrix of correlation
functions in which different superpositions of excited-state
contributions are linearly combined to isolate the energy eigenstates.
A low-lying Roper state was identified with this method using a
variety of source and sink smearings in constructing correlation
matrices~\cite{Mahbub:2009aa,Mahbub:2010jz} in quenched QCD. Recent
developments of algorithms and computational power have enabled the
extension to full QCD.  Some full QCD analyses using the variational
method can be found in
Refs.~\cite{Bulava:2009jb,Lin:2008pr,Engel:2010my,Edwards:2011jj,Bulava:2010yg,Mahbub:2010rm,Morningstar:2011ka,Bulava:2011uk,Menadue:2011pd,Edwards:2012fx,Engel:2013ig}.
Here we consider the techniques of
Refs.~\cite{Mahbub:2009aa,Mahbub:2010jz} to explore the low-lying
even- and odd- parity states of the nucleon using 2+1-flavor dynamical
QCD gauge-field configurations from the PACS-CS
collaboration~\cite{Aoki:2008sm}. A small subset of the results
presented here have appeared in
Refs.~\cite{Mahbub:2010rm,Mahbub:2012ri}.

The number of energy states revealed in the correlation matrix method
depends on the number of unique operators chosen which have the quantum
numbers of the desired states. Hence, a clear identification of these states
is necessary to observe changes in these energy-states as a function
of quark mass, in principle to the physical quark mass. This allows
the quark mass dependence and structure of the extracted energy
eigenstates to be explored systematically. The new technique that we
develop here can be used when any parameter of the theory is varied to
explore how the nature of the states and energies change with
that parameter.

The principal focus of this paper is to present the details of our
eigenvector analysis to track the states from the heavy to the light
quark mass region. In doing so, we consider the $N^{\pm}$ states to
illustrate the utility of the method. The operator basis is increased
with the use of fermion source and sink smearings as in
Ref.~\cite{Mahbub:2010rm}. Then, the propagation of the energy states
are presented after analyzing the state of eigenvectors at adjacent
quarks masses. Results are presented for both the non-symmetric and
symmetric eigenvalue equations, and these are compared and related to
each other.

In this analysis, we haven't been able to isolate the multi-particle
thresholds at our three light quark masses. We proceed under the
assumption that the couplings to these 5+ quark states have relatively
small overlap with our 3-quark interpolators and that the effective
mass functions are largely unaffected by multi-particle states with
small couplings to our interpolators. To monitor this we use the full
covariance-matrix based $\chi^2/\rm{dof}$ in order to accurately
assess the extent to which our effective mass function plateau is
associated with a single state.  Ultimately, in addition to the
3-quark operators, one needs to include 5- or 7- or more- quark
operators~\cite{Morningstar:2011ka,Lang:2012db,Morningstar:2013bda} in
the correlation matrix to extract all the states in the spectrum.

If we denote the dimensionality of the Hilbert space of the lattice
Hamiltonian to be $N$, then in an ideal world we would select $N$
linearly independent operators to construct our $N\times N$
correlation matrix with sufficient statistical accuracy and then
diagonalize this to obtain the exact $N$ energy eigenstates for this
lattice Hamiltonian. Obviously and unfortunately this is not
computationally feasible on any realistic lattice and the best that we
can do is choose a relatively small number of operators, $M$, where
$M<<N$. If we choose these $M$ operator interpolating fields wisely,
then the subspace they span will have good overlap with the subspace
spanned by the $M$ lowest eigenstates of the Hamiltonian. If this is
the case then with sufficient statistics we can hope to extract good
estimates of the $M$ lowest energy states. If we observe, for example,
that adding additional operators to increase $M$ to $M^{\prime}$
reveals new low-lying states, then clearly we had not chosen our $M$
operators wisely enough. The test of whether or not we have revealed
all of the lowest states of the lattice Hamiltonian is that the
process of adding additional and new operators does not reveal new
low-lying excited states. We present a clearer and more complete
discussion of these issues in the Appendix.

The linear independence of additional operators can be judged by
monitoring the condition number of the correlation matrix. If the
condition number does not increase significantly when an operator is
added then the additional operator enhances the basis in a
sufficiently independent manner.

The paper is arranged as follows:
Section~\ref{section:mass_of_hadrons} contains a standard description
of the mass extraction from a two-point correlation function with a
brief introduction of Gaussian smearings at the fermion sources. The
variational method is presented in
section~\ref{sec:variational_method} followed by simulation details in
section~\ref{section:simulation_details}. Section~\ref{sec:Eigenstate_Identification}
contains a discussion of the energy eigenstates identification.
Results for the flow of eigenvectors are presented in
section~\ref{sec:quark_mass_flow} followed by concluding remarks in
section~\ref{section:conclusions}. Finally, a pedagogical discussion
of the correlation matrix is presented in Appendix in terms of the
lattice Hamiltonian.

\section{Energy States from Two-point Correlation Functions}
\label{section:mass_of_hadrons}

A two point correlation function can be written as
\begin{align}
 {G_{ij}(t,\vec p)}&=\sum_{\vec x}e^{-i{\vec p}.{\vec x}}\langle{\Omega}\vert T \{ \chi_i (x)\bar\chi_j(0)\} \vert{\Omega}\rangle, 
\label{sec:mass:first_eqn}
\end{align}
where the Dirac spin indices are implicit.  The operator
$\bar{\chi}_j(0)$ creates states from the vacuum at space-time point
$0$ and, following the evolution of the states in Euclidean time $t$,
the states are destroyed by the operator $\chi_{i}(x)$ at the point
$(\vec{x},t)$. $T$ indicates the time ordering of the operators.

The energy eigenstates of hadrons are extracted using operators
suitably chosen to have overlap with the desired states of interest.
If we consider a baryon state $B$, then a complete set of momentum
eigenstates provides,

\begin{align}
\sum_{B,{\vec p}^{\, \prime},s}\vert{B,{\vec p}^{\,
    \prime},s}\rangle\langle{B,{\vec p}^{\, \prime},s}\vert &=I,
\label{sec:mass:completeness_eqn}
\end{align}
where $B$ can also include multi-particle states that
the operator $\chi$ couples with. The substitution of
Eq.~(\ref{sec:mass:completeness_eqn}) into
Eq.~(\ref{sec:mass:first_eqn}) yields
\begin{align}
{G_{ij}(t,\vec p)} &=\sum_{\vec x}\sum_{B,{\vec p}^{\,
    \prime},s}e^{-i{\vec p}.{\vec x}} \nonumber \\ & \langle
{\Omega}\vert\chi_i(x)\vert{B,{\vec p}^{\,
    \prime},s}\rangle\langle{B,{\vec p}^{\, \prime},s}\vert\bar\chi_j(0)\vert {\Omega}\rangle.\label{sec:mass:MinkowskiCorrelationFunction_eqn1}
\end{align}
Using the translational operator, the operator $\chi_{i}(x)$ can be
expressed as
\begin{align}
\chi_{i}(x) &= e^{Ht}e^{-i\vec{P}\cdot\vec{x}}\chi_{i}(0)e^{i\vec{P}\cdot\vec{x}}e^{-Ht},
\end{align}
where $H$ is the lattice Hamiltonian and $\vec{P}$ is the momentum operator
whose eigenvalue is the total momentum $\vec{p}$ of the system. Inserting this into Eq.~(\ref{sec:mass:MinkowskiCorrelationFunction_eqn1}) we obtain
\begin{align}
&{G_{ij}(t,\vec p)} =\sum_{\vec x}\sum_{B,{\vec p}^{\, \prime},s}e^{-E_{B}t}e^{-i{\vec x}.({\vec p}-{\vec p}^{\, \prime})} \nonumber \\
&\hspace{0.12\textwidth}\langle{\Omega}\vert\chi_{i}(0)\vert{B,{\vec p}^{\, \prime},s}\rangle\langle{B,{\vec p}^{\, \prime},s}\vert\bar\chi_{j}(0)\vert {\Omega}\rangle \nonumber \\
&=\sum_{B,{\vec p}^{\,
    \prime},s}e^{-E_{B}t}\delta_{\vec p,{\vec p}^{\, \prime}}\langle
{\Omega}\vert\chi_{i}(0)\vert{B,{\vec p}^{\,
    \prime},s}\rangle\langle{B,{\vec p}^{\, \prime},s}\vert\bar\chi_{j}(0)\vert {\Omega}\rangle\nonumber \\
&=\sum_{B}\sum_{s}e^{-E_{B}t}\langle {\Omega}\vert\chi_{i}(0)\vert{B,{\vec p},s}\rangle\langle{B,{\vec p},s}\vert\bar\chi_{j}(0)\vert {\Omega}\rangle.
\label{sec:mass:EucleadianCorrelationFunction_eqn1}
\end{align}
The overlap of the interpolating fields $\chi(0)$ and ${\bar\chi}(0)$
with positive and negative parity baryon states $\vert
{B^{\pm}}\rangle$ can be parametrized by a complex quantity called the
coupling strength, $\lambda_{B^{\pm}}$, which can be defined for
positive parity states by
\begin{align}
\langle{\Omega}\vert\chi(0)\vert {B^{+}},\vec p,s\rangle &=\lambda_{B^{+}}\sqrt {\frac{M_{B^{+}}}{E_{B^{+}}}}u_{B^{+}}({\vec p},s),
\end{align}
\begin{align}
\langle B^{+},\vec p,s\vert\bar{\chi}(0)\vert {\Omega}\rangle &=\bar\lambda_{B^{+}}\sqrt {\frac{M_{B^{+}}}{E_{B^{+}}}}{{\bar u}_{B^{+}}}({\vec p},s).
 \label{eq:def_lambdabar_posp}
\end{align}
For the negative parity states one requires

\begin{align}
\langle{\Omega}\vert\chi(0)\vert {B^{-}},\vec p,s\rangle &=\lambda_{B^{-}}\sqrt {\frac{M_{B^{-}}}{E_{B^{-}}}}\gamma_5{u_{B^{-}}({\vec p},s)},
\end{align}

\begin{align}
\langle B^{-},\vec p,s\vert\bar{\chi}(0)\vert {\Omega}\rangle &=-\bar\lambda_{B^{-}}\sqrt {\frac{M_{B^{-}}}{E_{B^{-}}}}{{\bar u}_{B^{-}}}({\vec p},s)\gamma_5.
\end{align}
Here, $\lambda_{B^{\pm}} $ and ${\bar\lambda}_{B^{\pm}}$ are the
couplings of the interpolating functions at the sink and the source
respectively and $M_{B^{\pm}}$ is the mass of the state
$B^{\pm}$. ${E_{B^{\pm}}}$ is the energy of the state $B^{\pm}$, where
${E_{B^{\pm}}} = \sqrt{M^{2}_{B^{\pm}}+{\vec p}^2}$. Therefore, mass
of a energy-state is obtained with the momentum projection of the
correlation function at $\vec{p}=0$. 

The standard spin sums may now be performed. For the
positive parity hadron states, this can be expressed as
\begin{align} 
\sum_{s} {u^{\beta}_{B^{+}}}(\vec p,s){\bar u}^{\alpha}_{B^{+}}(\vec p,s) &=\frac{\gamma .p + M_{B^{+}}}{2{M_{B^{+}}}},
\end{align}
and for the negative parity states, one encounters
\begin{align} 
- \gamma_{5}\left(\sum_{s} {u^{\beta}_{B^{-}}}(\vec p,s){\bar u}^{\alpha}_{B^{-}}(\vec p,s)\right)\gamma_{5} &=\frac{+\gamma .p - M_{B^{-}}}{2{M_{B^{-}}}}.
\end{align}
By substituting the above Eqs. for the positive and negative parity states in Eq.~(\ref{sec:mass:EucleadianCorrelationFunction_eqn1}) we obtain,
\begin{align}
{\cal{G}}_{ij}(t,\vec p) &=\sum_{B^{+}}\lambda_{B^{+}}\bar\lambda_{B^{+}}e^{{-E_{B^{+}}}t} {\frac{\gamma .p_{B^{+}} + M_{B^{+}}}{2E_{B^{+}}}} \nonumber \\
              & +\sum_{B^{-}}\lambda_{B^{-}}\bar\lambda_{B^{-}}e^{{-E_{B^{-}}}t} {\frac{+\gamma .p_{B^{-}} - M_{B^{-}}}{2E_{B^{-}}}}.\label{sec:mass:FinalCorrelationFunction_eqn}
\end{align}
At momentum $\vec p=\vec 0$, $E_{B^{\pm}}=M_{B^{\pm}}$, and a parity projection operator $\Gamma_{\pm}$ can be introduced,
\begin{align}
\Gamma_{\pm} &= \frac{1}{2}(\gamma_0 \pm 1).
\end{align}
 We can isolate the masses of the even and odd parity energy-states by taking
 the trace of $\cal{G}$ with the operators $\Gamma_{+}$ and
 $\Gamma_{-}$. The positive parity states propagate through the $(1,1)$
 and $(2,2)$ elements of the Dirac matrix, whereas, negative parity
 states propagate through the $(3,3)$ and $(4,4)$ elements.
The correlation function for positive and negative parity states can then be written as
\begin{align}
G_{ij}^{\pm}(t,\vec 0) &= {\rm{Tr}}_{\rm sp}[\Gamma_{\pm}{\cal{G}}_{ij}(t,\vec 0)]\nonumber \\
&= \sum_{B^{\pm}}\lambda_{i}^{\pm}\bar\lambda_{j}^{\pm}e^{{-M_{B^{\pm}}}t}.
\end{align}
The correlation function contains a superposition of energy-states,
i.e. both ground and excited energy-states. The mass
of the lowest energy-state, $M_{0^{\pm}}$ can be extracted at large $t$ where
the contributions from all other excited-states are suppressed,
\begin{align}
G_{ij}^{\pm}(t,\vec 0) & \stackrel{t \rightarrow \infty}{=} \lambda_{i0}^{\pm}\bar\lambda_{j0}^{\pm}e^{{-M_{0^{\pm}}}t},
\label{eq:lowest_neergy_state}
\end{align}
where, $\lambda_{i0}^{\pm}$ and $\bar\lambda_{j0}^{\pm}$ are now
couplings of interpolators to the lowest energy-state.

\subsection{Source Smearing}
The spatial fermion source-smearing~\cite{Gusken:1989qx} technique is
applied to increase the overlap of the interpolators with the lower
lying states. We employ a fixed boundary condition in the time
direction for the fermions by setting $U_t(\vec
x,N_t)=0\,\forall\,{\vec x}$ in the hopping terms of the fermion
action with periodic boundary conditions imposed in the spatial
directions. Gauge invariant Gaussian smearing ~\cite{Gusken:1989qx} in
the spatial dimensions is applied through an iterative process. The
smearing procedure is:
\begin{align}
\psi_{i}(\vec{x},t) &=\sum_{x'}F(\vec{x},\vec{x}')\psi_{i-1}(\vec{x}',t),
\end{align}
where,
\begin{align}
F(\vec{x},\vec{x}') &= {(1-\alpha)}\delta_{x,x'}+\frac{\alpha}{6}\sum_{\mu=1}^{3}[U_{\mu}(x)\delta_{x',x+\hat\mu} \nonumber \\
        & +U_{\mu}^{\dagger}(x-\hat\mu)\delta_{x',x-\hat\mu}],
\end{align}
where the parameter $\alpha=0.7$ is used in our calculation. After
repeating the procedures $N_{\rm sm}$ times on a point source the
resulting smeared fermion field is,
\begin{align}
\psi_{N_{\rm sm}}(\vec{x},t) &=\sum_{x'}F^{N_{sm}}(\vec{x},\vec{x}')\psi_{0}(\vec{x}',t).
\end{align}

\section{Variational Method}
 \label{sec:variational_method}

The extraction of the ground state mass can be done straightforwardly
using Eq.~(\ref{eq:lowest_neergy_state}). However access to the
excited state masses requires additional effort due to the presence of
these energy-states at the sub-leading of the exponential. Here we
consider the variational method ~\cite{Michael:1985ne,Luscher:1990ck},
which allows for a variety of superpositions of excited-states in its
cross-correlation discussed below.

The variational
method requires the cross correlation of operators so that the
operator space can be diagonalised and the excited state masses
extracted from the exponential nature of the diagonalised basis. To
access $N$ states of the spectrum, one requires a minimum of $N$
interpolators. With the assumption that only $N$ states contribute
significantly to $G_{ij}$ at time $t$, the parity projected two point
correlation function matrix for $\vec{p} =0$ can be written as 
\begin{align}
G^{\pm}_{ij}(t) & \equiv G^{\pm}_{ij}(t,\vec{0}) \nonumber \\ 
 &= (\sum_{\vec x}{\rm Tr}_{\rm sp}\{ \Gamma_{\pm}\langle\Omega\vert\chi_{i}(x)\bar\chi_{j}(0)\vert\Omega\rangle\}) \nonumber \\
 &=\sum_{\alpha=0}^{N-1}\lambda_{i}^{\alpha}\bar\lambda_{j}^{\alpha}e^{-m_{\alpha}t},
 \label{eq:cor_at_zero_mom}
\end{align}
where Dirac indices are implicit. Here, $\lambda_{i}^{\alpha}$ and
$\bar\lambda_{j}^{\alpha}$ are the couplings of interpolators
$\chi_{i}$ and $\bar\chi_{j}$ at the sink and source respectively to
eigenstates $\alpha=0, \cdots ,(N-1)$ and $m_{\alpha}$ is the mass of
the energy-state $\alpha$. The use of identical source and sink
interpolators provides
$\bar\lambda_{j}^{\alpha}=(\lambda_{j}^{\alpha})^{\ast}$ and then in
the ensemble average $G^{\pm}_{ij}(t)$ is a Hermitian matrix,
i.e. $G^{\pm}_{ij}(t)=[G^{\pm}_{ji}(t)]^{\ast}$. Moreover, considering
both $\{U\}$ and $\{U^{\ast}\}$ configurations makes $G^{\pm}_{ij}(t)$
a real symmetric matrix. The $N$ interpolators have the same quantum
numbers and provide an $N$-dimensional basis upon which to describe
the states. Using this basis we aim to construct $N$ independent
interpolating source and sink fields which isolate $N$ baryon states
$\vert B_{\alpha}\rangle,$ {\it i.e.}
\begin{align}
{\bar\phi}^{\alpha} &=\sum_{i=1}^{N}u_{i}^{\alpha}{\bar\chi}_{i},
 \label{eq:def_bar_phi_alpha}
\end{align}   
\vspace{-0.60cm}
\begin{align}
{\phi}^{\alpha} &=\sum_{i=1}^{N}v_{i}^{\alpha}{\chi}_{i},
\end{align} 
such that,
\begin{align}
\langle{B_{\beta},p,s}\vert {\bar\phi}^{\alpha}\vert\Omega\rangle &= \delta_{\alpha\beta}{\bar{z}}^{\alpha}\bar{u}(\alpha,p,s),
\end{align}
\vspace{-0.60cm}
\begin{align}
\langle\Omega\vert{\phi}^{\alpha}\vert B_{\beta},p,s\rangle &= \delta_{\alpha\beta}{z}^{\alpha}u(\alpha,p,s),
\end{align}
where $z^{\alpha}$ and ${\bar{z}}^{\alpha}$ are the coupling strengths
of $\phi^{\alpha}$ and ${\bar\phi}^{\alpha}$ to the state $\vert
B_{\alpha}\rangle$. Consider a real eigenvector $u_{j}^{\alpha}$ which
operates on the correlation matrix $G_{ij}(t)$ from the right, one can
obtain, 
\begin{align}
G_{ij}(t)u_{j}^{\alpha} &=(\sum_{\vec x}{\rm Tr}_{\rm sp}\{ \Gamma_{\pm}\langle\Omega\vert\chi_{i}\bar\chi_{j}\vert\Omega\rangle\})u_{j}^{\alpha} \nonumber \\
& = \lambda_{i}^{\alpha}\bar{z}^{\alpha}e^{-m_{\alpha}t}.
\label{eq:cmmatrix_first}
\end{align}
For notational convenience, in the remainder of the discussion the
repeated indices $i,j,k$ are to be understood as being summed over,
whereas, $\alpha$, which stands for a particular state, is not.

In the ensemble average, $G_{ij}(t)=G_{ji}(t)$. Therefore, considering
$\frac{1}{2}[G_{ij}(t)+G_{ji}(t)]$ provides an improved unbiased
estimator and enables the use a symmetric eigenvalue equation as
discussed below. To ensure that the matrix elements are all
$\sim{\cal{O}}(1)$, each element of $G_{ij}(t)$ is normalized by
$\frac{1}{\sqrt{{G}_{ii}(0)}}{G}_{ij}(t)\frac{1}{\sqrt{{G}_{jj}(0)}}$
(discussed in the Appendix).

 In Eq.~(\ref{eq:cmmatrix_first}), since the only $t$ dependence comes
 from the exponential term, we can write a recurrence relation at time
 $(t_{0}+\triangle t)$ as,
\begin{align}
G_{ij}(t_{0}+\triangle t)u_{j}^{\alpha} & = e^{-m_{\alpha}\triangle t} G_{ij}(t_{0})u_{j}^{\alpha},
\end{align}  
for sufficiently large $t_{0}$ and $t_{0}+\triangle t$~\cite{Blossier:2009kd,Mahbub:2009nr}.

Multiplying the above equation by $[G_{ij}(t_{0})]^{-1}$ from the left we get,
\begin{align}
[(G(t_{0}))^{-1}G(t_{0}+\triangle t)]u^{\alpha} & = e^{-m_{\alpha}\triangle t}u^{\alpha} \nonumber \\
& = c^{\alpha}u^{\alpha}.
\label{eqn:right_eigenvalue_equation}
\end{align} 
This is an eigenvalue equation for eigenvector $u^{\alpha}$ with
eigenvalue $c^{\alpha}=e^{-m_{\alpha}\triangle t}$. We can also solve
the left eigenvalue equation to recover the $v^{\alpha}$ eigenvector,
\begin{align}
v_{i}^{\alpha}G_{ij}(t_{0}+\triangle t) & = e^{-m_{\alpha}\triangle t}v_{i}^{\alpha}G_{ij}(t_{0}).
\end{align} 
Similarly,
\begin{align}
v^{\alpha}[G(t_{0}+\triangle t)(G(t_{0}))^{-1}] & = e^{-m_{\alpha}\triangle t}v^{\alpha}.
\label{eqn:left_eigenvalue_equation}
\end{align} 

Since $G_{ij}(t)$ is a real symmetric matrix $v=u$. The vectors
$u_{j}^{\alpha}$ and $v_{i}^{\alpha}$ diagonalize the correlation
matrix at time $t_{0}$ and $t_{0}+\triangle t$ making the projected
correlation matrix,
\begin{align}
v_{i}^{\alpha}G_{ij}(t)u_{j}^{\beta} = \delta^{\alpha\beta}z^{\alpha}{\bar{z}}^{\beta}e^{-m_{\alpha}t}.
 \label{eqn:projected_cf} 
\end{align} 
The parity projected, eigenstate projected correlator,
\begin{align}
v_{i}^{\alpha}G^{\pm}_{ij}(t)u_{j}^{\alpha} \equiv G_{\pm}^{\alpha}
\end{align}
is then used to obtain masses of different states.
We construct the effective mass function
\begin{align}
M_{\rm eff}^{\alpha}(t) &= {\rm ln}\left(\frac{{G_{\pm}^{\alpha}}(t,\vec 0)}{G_{\pm}^{\alpha}(t+1,\vec 0)}\right)\nonumber \\
 & = M_{\pm}^{\alpha}.
 \label{eqn:efective_mass}
\end{align}
and apply standard analysis techniques as described in Ref.~\cite{Mahbub:2009nr}.

Since ${G(t_{0})}^{-1/2}{G(t_{0})}^{1/2}=I$, we can rewrite Eq.~(\ref{eqn:right_eigenvalue_equation})as
\begin{align}
G(t_{0})^{-1}G(t_{0}+\triangle t){G(t_{0})}^{-1/2}{G(t_{0})}^{1/2}u^{\alpha} & = c^{\alpha}u^{\alpha} \nonumber 
\end{align} 
Multiplying from the left by ${G(t_{0})}^{1/2}$ provides
\begin{align}
{G(t_{0})}^{-1/2}G(t_{0}+\triangle t){G(t_{0})}^{-1/2}{G(t_{0})}^{1/2}u^{\alpha} & = c^{\alpha}{G(t_{0})}^{1/2}u^{\alpha} \nonumber
\end{align}
and defining, \begin{align} w^{\alpha} = {G(t_{0})}^{1/2} u^{\alpha} \end{align} we find
\begin{align}
{G(t_{0})}^{-1/2}G(t_{0}+\triangle t){G(t_{0})}^{-1/2}w^{\alpha} & = c^{\alpha}w^{\alpha}
 \label{eqn:symmetric_evalue}
\end{align} 
(also shown in Eq.~(\ref{eq:sqrtgt0_gt0plusdeltat_sqrtgt0})). We note
the matrix
\begin{align}[{G(t_{0})}^{-1/2}G(t_{0}+\triangle t){G(t_{0})}^{-1/2}] \label{eqn:symmetric_matrix} \end{align} 
is real symmetric, with the same eigenvalue $c^{\alpha}$ and with the
$\vec{w}^{\alpha}$ orthogonal to each other. If we had not used the
$[U+U^{\ast}]$ sum then the matrix in
Eq.~(\ref{eqn:symmetric_matrix}) would be hermitian and hence would
still have real eigenvalues and orthogonal eigenvectors.  The
coefficients of interpolators creating an energy eigenstate is
recovered by
\begin{align} u^{\alpha} = {G(t_{0})}^{-1/2} w^{\alpha} \, . \end{align} 
\section{Simulation parameters}
\label{section:simulation_details}
PACS-CS $2+1$ flavor dynamical-fermion
configurations~\cite{Aoki:2008sm} made available through the ILDG
\cite{Beckett:2009cb} are used. These configurations use the
non-perturbatively ${\cal{O}}(a)$-improved Wilson fermion action and
the Iwasaki-gauge action~\cite{Iwasaki:1983ck}. The lattice volume is
$32^{3}\times 64$, with $\beta=1.90$ providing a lattice spacing
$a=0.0907$ fm and lattice volume of $\approx (2.90\, \rm{fm})^{3}$.

The degenerate up and down quark masses are considered, with the
hopping parameter values of $\kappa_{ud}=0.13700, 0.13727, 0.13754,
0.13770\text{ and }0.13781$ corresponding to pion masses of $m_{\pi}$
= 0.702, 0.572, 0.413, 0.293, 0.156 GeV~\cite{Aoki:2008sm}; for the
strange quark $\kappa_{s}=0.13640$. We consider an ensemble of 350
configurations each for the four heavier quarks mass and 198
configurations for the lightest quark. An ensemble of 750 samples for
the lightest quark mass is created by using well separated multiple fermion sources
on each configuration.  We use the jackknife method to calculate
the error, where the ${\chi^{2}}/{\rm{dof}}$ for projected correlator fits
is obtained via a covariance matrix analysis.

The complete set of local interpolating fields for the
spin-$\frac{1}{2}$ nucleon are considered herein. Three different
spin-flavor combinations of nucleon interpolators are considered, 
\begin{align}
\label{eq:interp_x1}
\chi_1(x) &= \epsilon^{abc}(u^{Ta}(x)C{\gamma_5}d^b(x))u^{c}(x)\, , \\
\label{eq:interp_x2}
\chi_2(x) &= \epsilon^{abc}(u^{Ta}(x)Cd^b(x)){\gamma_5}u^{c}(x)\, , \\
\label{eq:interp_x4}
\chi_4(x) &= \epsilon^{abc}(u^{Ta}(x)C{\gamma_5}{\gamma_4}d^b(x))u^{c}(x).
\end{align}
The $\chi_{1}$ and $\chi_{2}$ interpolators are used in
Refs.~\cite{Leinweber:1990dv,Sasaki:2001nf,Leinweber:1994nm}.  The
interpolator $\chi_{4}$ is the time component of the local
spin-$\frac{3}{2}$ isospin-$\frac{1}{2}$ interpolator which also
couples to spin-$\frac{1}{2}$ states used, for instance, in
Refs.~\cite{Brommel:2003jm,Zanotti:2003fx,Mahbub:2009nr}.

The local scalar-diquark nucleon interpolator, $\chi_{1}$, is well
known to have a good overlap with the ground state of the
nucleon. Also, this $\chi_{1}$ interpolator is able to extract a
low-lying Roper state in quenched QCD~\cite{Mahbub:2010jz}. On the
other hand, the $\chi_{2}$ interpolator has pseudoscalar-diquark
structure in the nucleon, which vanishes in the non-relativistic
limit, couples strongly to higher energy states.  Each interpolator
has a unique Dirac structure giving rise to different spin-flavor
combinations. Moreover, as each spinor has upper and lower components,
with the lower components containing an implicit derivative, different
combinations of zero, one, two and three derivative interpolators are provided.

The correlation matrices are constructed using different levels of
gauge-invariant Gaussian smearing \cite{Gusken:1989qx} at the fermion
sources and sinks~\cite{Mahbub:2010rm}. A basis of smearing-sweep
counts of 16, 35, 100 and 200 is selected following the extensive
analysis of Ref.~\cite{Mahbub:2010rm}.

It is important to consider the condition number for these matrices in
order to examine the quality of our operator basis.  We consider the
normalized correlation matrix,
$G_{ij}(t)/(G_{ii}(t)G_{jj}(t))^{-1/2}$, with $G_{ij}(t)$ made
Hermition as discussed in Sec.~\ref{sec:variational_method}.

\begin{figure}[!h]
  \begin{center}
 \includegraphics[height=0.47\textwidth,angle=90]{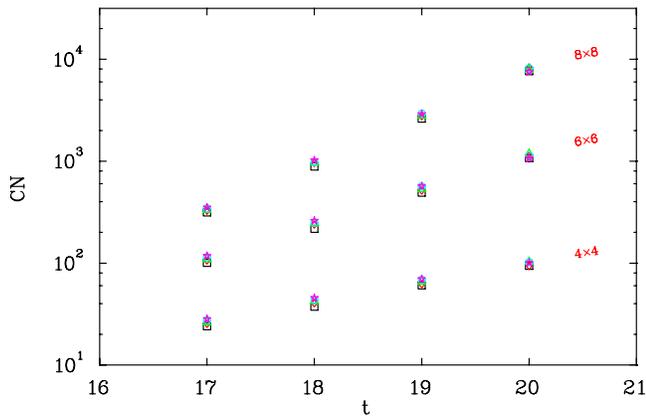}
    \caption{(Color online) The condition numbers, CN,
        of $4\times 4$, $6\times 6$ and $8\times 8$ correlation
        matrices using $\chi_{1}$ and $\chi_{2}$ operators, are
        illustrated as a function of Euclidean time, $t$.  The
        $4\times 4$ matrix includes 200 and 100 sweeps, $6\times 6$
        contains 200, 100 and 35 sweeps and $8\times 8$ incorporates
        all four sources, 200, 100, 35 and 16 sweeps. Each cluster of
        points contains five values corresponding to the five quark
        masses considered.}
  \label{fig:ConditionNumbers.4x4.6x6.8x8}
  \end{center} 
\end{figure}

The condition numbers for our correlation matrices are illustrated in
Fig.~\ref{fig:ConditionNumbers.4x4.6x6.8x8}.  We examine the change in
the condition number for matrices composed of $\chi_1$ and $\chi_2$ as
additional source smearings are introduced.  We consider two levels of
smearing in the $4 \times 4$ matrix, three levels of smearing in the
$6 \times 6$ matrix and all four levels of smearing in the $8 \times
8$ matrix.

Results for the five quark masses under consideration are provided.
The tight clustering of the condition number for the wide range of
quark masses considered indicates that the basis selected is
appropriate for all these masses.

The condition numbers are displayed as a function of Euclidean time
following the fermion source at $t_s=16$.  At early times, the
different superposition of large excited state contributions gives
rise to a relatively small condition number.  However as these states
become exponentially suppressed at larger Euclidean times, the
condition number increases.  If one waits to very large Euclidean
times, all excitations are suppressed and all operators produce the
same ground state rendering the condition number infinite.  This will
be realised for any basis set having overlap with the ground state.
Thus it is important to conduct the correlation matrix analysis at
times where excited states are present and the number of significant
state contributions matches the size of the basis.

While the condition number increases as the smearing basis is
enhanced, the condition number is the order of $10^3$ for our
preferred variational analysis time of $t_0 = 18$.  This value is
small relative to $10^{12}$ associated with standard double precision
calculations.

Thus, the utilization of different fermion smearings at the source and
the sink is an effective approach to enlarging the basis of operators.
Our selection of smearing levels was based on the excited-state
contributions observed in smeared-source to point-sink
correlators~\cite{Mahbub:2010rm}. By selecting smearing levels that
provided well separated effective masses at early Euclidean times, we
ensured that each operator was sufficiently independent, thus giving
rise to an acceptable condition number for the correlation matrix.

\section{Eigenstates Identification}
 \label{sec:Eigenstate_Identification}

Let us consider $M$ interpolating fields making an $M\times M$
parity-projected correlation matrix $G(t)$.  In solving the
generalized eigenvalue equations of
Eqs.~(\ref{eqn:right_eigenvalue_equation}) and
(\ref{eqn:left_eigenvalue_equation}) we encounter the real and
approximately symmetric matrices $[(G(t_{0}))^{-1}\, G(t_{0}+\triangle
  t)]$ and $[G(t_{0}+\triangle t)\, (G(t_{0}))^{-1}]$, with the left
and right eigenvectors $\vec{u}^{\alpha}$ and $\vec{v}^{\alpha}$
respectively.  Thus the eigenvectors of these matrices are expected to
be approximately orthogonal (left table in
Table~\ref{table:udotu_wdotw_heaviestkappa_x1x2}). As explained in the
Appendix, the reason we have only approximate symmetry is that $G(t)$
does not commute with itself at different times. This results because
$M<N$. The more closely the subspace spanned by our $M$ operators
aligns with the subspace of the lowest $M$ energy eigenstates of $H$,
the less violation of symmetry there will be. If we do not use the
$[U+U^{\ast}]$ sum, then all of the same arguments hold but with
hermitian matrices.

This feature enables the use of the generalised measure
\begin{align}
{\mathcal U}^{\alpha \beta}(m_q, m_{q^\prime}) & =
\vec{u}^{\alpha}(m_{q}) \cdot \vec{u}^{\beta}(m_{q^\prime})
 \label{eq:generalised_measure_u}
\end{align}
for the eigenvector $\vec{u}^{\alpha}$, for example. This correlates
eigenvectors at different quark masses and may be useful in tracking
states.

In contrast, as already discussed, the matrix in
Eq.~(\ref{eqn:symmetric_matrix}) is symmetric, hence
the eigenvectors $\vec{w}^{\alpha}(m_{q})$ are exactly orthogonal,
i.e. $\vec{w}^{\alpha}(m_{q})\cdot
\vec{w}^{\beta}(m_{q})=\delta_{\alpha\beta}$ (right table in Table~\ref{table:udotu_wdotw_heaviestkappa_x1x2}).

\begin{table*}
 \begin{center}
 \caption{The scalar product $\vec{u}^{\alpha}(m_{q}) \cdot
   \vec{u}^{\beta}(m_{q})$ (left) and $\vec{w}^{\alpha}(m_{q}) \cdot
   \vec{w}^{\beta}(m_{q})$ (right), for the same quark mass, with four
   different levels of smearings.  States are ordered from left to
   right and top to bottom in order of increasing excited-state
   mass. $\alpha$ and $\beta$ correspond to row and column,
   respectively.}
 \label{table:udotu_wdotw_heaviestkappa_x1x2}
 \vspace{0.3cm}
 \begin{tabular}{cccccccc||cccccccc}
 \hline
 \hline                                        
  \textbf{1.00} & -0.18 & 0.02 & -0.07 & 0.65 & 0.10 & -0.32 & -0.09   &    \textbf{1.00} & 0.00 & 0.00 & 0.00 & 0.00 & 0.00 & 0.00 & 0.00 \\  
  -0.18 & \textbf{1.00} & 0.02 & 0.36 & -0.10 & -0.49 & 0.06 & 0.39    &    0.00 & \textbf{1.00} & 0.00 & 0.00 & 0.00 & 0.00 & 0.00 & 0.00 \\  
  0.02 & 0.02 & \textbf{1.00} & 0.15 & 0.07 & 0.06 & 0.42 & 0.03       &    0.00 & 0.00 & \textbf{1.00} & 0.00 & 0.00 & 0.00 & 0.00 & 0.00 \\     
  -0.07 & 0.36 & 0.15 & \textbf{1.00} & -0.03 & 0.23 & 0.09 & 0.30     &    0.00 & 0.00 & 0.00 & \textbf{1.00} & 0.00 & 0.00 & 0.00 & 0.00 \\   
  0.65 & -0.10 & 0.07 & -0.03 & \textbf{1.00} & 0.15 & -0.57 & -0.13   &    0.00 & 0.00 & 0.00 & 0.00 & \textbf{1.00} & 0.00 & 0.00 & 0.00 \\  
  0.10 & -0.49 & 0.06 & 0.23 & 0.15 & \textbf{1.00} & -0.06 & -0.61    &    0.00 & 0.00 & 0.00 & 0.00 & 0.00 & \textbf{1.00} & 0.00 & 0.00 \\  
  -0.32 & 0.06 & 0.42 & 0.09 & -0.57 & -0.06 & \textbf{1.00} & 0.17    &    0.00 & 0.00 & 0.00 & 0.00 & 0.00 & 0.00 & \textbf{1.00} & 0.00 \\  
  -0.09 & 0.39 & 0.03 & 0.30 & -0.13 & -0.61 & 0.17 & \textbf{1.00}    &    0.00 & 0.00 & 0.00 & 0.00 & 0.00 & 0.00 & 0.00 & \textbf{1.00} \\  
  \hline
\end{tabular}
\end{center}
\end{table*}

Therefore, as in Eq.~(\ref{eq:generalised_measure_u}), a generalised measure
\begin{align}
{\mathcal W}^{\alpha \beta}(m_q, m_{q^\prime}) & =
\vec{w}^{\alpha}(m_{q}) \cdot \vec{w}^{\beta}(m_{q^\prime})
 \label{generalised_measure}
\end{align}
for the $\vec{w}^{\alpha}$ can be constructed to identify the states
more reliably as we move from quark mass $m_{q}$ to the adjacent
quark mass $m_{q^\prime}$.

\begin{table*}
 \begin{center}
 \caption{The scalar product $\vec{u}^{\alpha}(m_{q}) \cdot
   \vec{u}^{\beta}(m_{q^{\prime}})$ (left) and
   $\vec{w}^{\alpha}(m_{q}) \cdot \vec{w}^{\beta}(m_{q^{\prime}})$
   (right), for $\kappa=0.13700$ and $\kappa^{\prime}=0.13727$, with
   four different levels of smearings.  States are ordered from left
   to right for $m_{q^{\prime}}$ and top to bottom for $m_{q}$ in
   order of increasing excited-state mass. $\alpha$ and $\beta$
   correspond to row and column, respectively.}
 \label{table:udotu_wdotw_heaviest_second_heaviest_k_x1x2}
 \vspace{0.3cm}
 \begin{tabular}{cccccccc||cccccccc}
 \hline
 \hline                                        
  \textbf{0.98} & -0.29 & -0.14 & 0.63 & -0.07 & 0.10 & -0.32 & -0.08    &   \textbf{1.00} & -0.09 & 0.00 & 0.00 & 0.01 & 0.00 & 0.01 & 0.00     \\
  -0.19 & \textbf{-0.92} & 0.08 & -0.03 & 0.14 & 0.06 & 0.42 & 0.05      &   0.09 & \textbf{0.99} & -0.07 & 0.13 & -0.01 & 0.00 & 0.01 & 0.00    \\ 
  -0.16 & 0.07 & \textbf{0.99} & -0.09 & -0.04 & -0.53 & 0.09 & 0.36     &   0.01 & 0.07 & \textbf{1.00} & -0.01 & 0.00 & -0.01 & 0.00 & 0.00    \\
  0.63 & -0.44 & -0.02 & \textbf{0.99} & -0.05 & 0.13 & -0.55 & -0.12    &   -0.01 & -0.13 & 0.02 & \textbf{0.98} & -0.09 & 0.02 & 0.07 & 0.00   \\
  -0.12 & -0.11 & 0.40 & 0.00 & \textbf{0.75} & 0.00 & 0.08 & 0.36       &   0.01 & 0.01 & 0.00 & -0.09 & \textbf{-0.97} & 0.21 & -0.01 & 0.03   \\  
  0.05 & -0.11 & -0.42 & 0.17 & 0.76 & \textbf{0.95} & -0.12 & -0.53     &   0.00 & 0.00 & 0.01 & 0.00 & 0.20 & \textbf{0.95} & -0.07 & -0.23    \\
  -0.45 & -0.17 & 0.03 & -0.67 & 0.08 & -0.05 & \textbf{1.00} & 0.18     &   -0.01 & 0.00 & 0.00 & -0.07 & 0.01 & 0.07 & \textbf{0.99} & -0.01   \\
  -0.09 & 0.00 & 0.34 & -0.14 & -0.34 & -0.82 & 0.21 & \textbf{1.00}     &   0.00 & 0.00 & 0.00 & -0.01 & -0.08 & -0.21 & 0.01 & \textbf{-0.97}  \\
  \hline
\end{tabular}
\end{center}
\end{table*}

In Table~\ref{table:udotu_wdotw_heaviest_second_heaviest_k_x1x2}, the
generalised measures ${\mathcal U}^{\alpha \beta}(m_q, m_{q^\prime})$
and ${\mathcal W}^{\alpha \beta}(m_q, m_{q^\prime})$ are presented for
the heaviest and the second heaviest quark masses. It is evident that
the off-diagonal elements of ${\mathcal W}^{\alpha \beta}(m_q,
m_{q^\prime})$ are smaller than the ${\mathcal U}^{\alpha
  \beta}(m_q, m_{q^\prime})$ and hence will be more reliable for
identification and tracking of the energy eigenstates. Therefore we
use ${\mathcal W}^{\alpha \beta}(m_q, m_{q^\prime})$ for this
purpose. For each value of $\alpha$ there is only one value for
$\beta$ where the entry is within a few percent of 1.  Thus this
measure provides a clear identification of how eigenvectors in the
hadron spectrum at quark mass $m_q$ are associated with eigenvectors
at the next value of quark mass, $m_{q^\prime}$.

\begin{figure}[!h]
  \begin{center}
 \includegraphics[height=0.46\textwidth,angle=90]{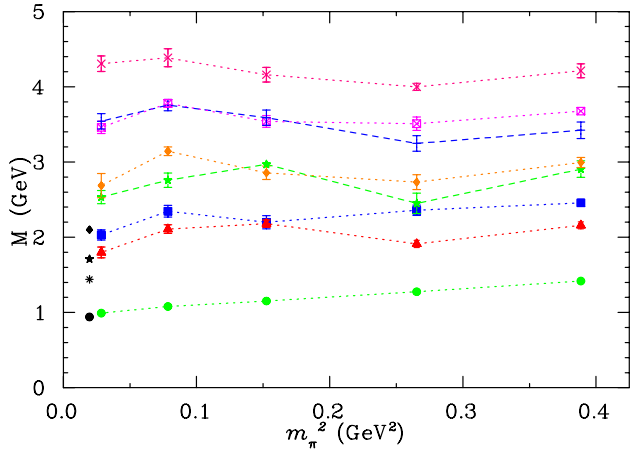}
    \caption{(Color online) $N{\frac{1}{2}}^{+}$ energy-states from
      $8\times 8$ correlation matrix of $\chi_{1},\chi_{2}$
      interpolators from $\kappa=0.13700$ ($m_{\pi}=702$ MeV, right-most
      column) to $\kappa=0.13781$ ($m_{\pi}=156$ MeV, left-most column).
      The symbols follow the eigenvector as determined by considering the scalar product
      $\vec{w}^{\alpha}\cdot \vec{w}^{\beta}$, as presented in
      Table~\ref{table:wdotw_forallkappa_x1x2}. Note that the dotted lines in
      the figure connect similar eigenvectors. Where these lines cross, of
      course the energy levels would not cross, but we would see an
      avoided level crossing as in Fig.~\ref{fig:avoided_Level_Crossing} if we had data for every quark or pion mass.}
  \label{fig:m.x1x2.8x8.forall_kappa}
  \end{center} 
\end{figure}

Now we explain how we track the energy eigenstates from one quark
mass to the next. Firstly, we label the extracted energy-states at the
heaviest quark mass with a chosen set of symbols (most right column in
Fig.~\ref{fig:m.x1x2.8x8.forall_kappa}), where each symbol is assigned
by the corresponding eigenvectors associated with it. These symbols
are carried on to the lightest quark mass by looking at ${\mathcal
  W}^{\alpha \beta}(m_q, m_{q^\prime})$ for adjacent quark masses going
from the heaviest to the lightest. After the energy eigenstates are
labeled at the heaviest quark, we look at the scalar product
$\vec{w}^{\alpha}(m_{q}) \cdot \vec{w}^{\beta}(m_{q^\prime})$ for the
heaviest and the second heaviest quark mass (top left of
Table~\ref{table:wdotw_forallkappa_x1x2}). The scalar product shows
that in this case all the diagonal elements are larger than the
off-diagonal, meaning there is no eigenvector crossing at these two
heavier quark masses. A similar scalar product for the second
heaviest to the third heaviest mass (top right of
Table~\ref{table:wdotw_forallkappa_x1x2}) shows that the fourth and
fifth eigenvectors are crossed with the fifth and fourth at the third
quark mass, and a similar situation for the sixth and the seventh. To
illustrate our analysis in our figures we track the eigenvectors from
one quark mass to the next by connecting these similar eigenvectors by
lines. In Fig.~\ref{fig:m.x1x2.8x8.forall_kappa}, two lines
(eigenvectors) cross at the second and the third quark mass. We then
follow the above procedures for the third, fourth and fourth and fifth
(the lightest) quark masses.

\begin{figure}[!h]
  \begin{center}
 \includegraphics[height=0.46\textwidth,angle=90]{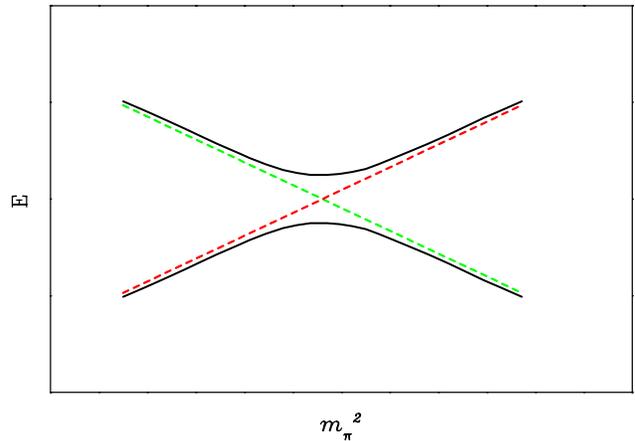}
    \caption{(Color online) Illustration of an avoided level
      crossing. The solid lines illustrate how the energy levels avoid
      crossing, while the two dotted lines illustrate how the nature
      of the associated eigenvectors flow. In the region where the
      energy states are closest each is an equal but orthogonal
      admixture of the two eigenvectors.}
  \label{fig:avoided_Level_Crossing}
  \end{center} 
\end{figure}

It is well known in quantum mechanics the energies avoid level
crossings as illustrated in
Fig.~\ref{fig:avoided_Level_Crossing}. However, when two energy levels
experience an avoided level crossing, the nature of the two
eigenvectors is interchanged, as shown by the dotted lines in
Fig.~\ref{fig:avoided_Level_Crossing}. In
Fig.~\ref{fig:m.x1x2.8x8.forall_kappa}, the first and second excited
energy eigenstates do not experience an avoided level crossing at pion
mass of $413$ MeV, whereas avoided level crossings are present for the
third-fourth and the fifth-sixth excited energy-states. However, note
that the avoided level crossings lie within the error bars.  Results
are presented as a function of quark mass $(m_{q})$, with
$m_{q^{\prime}}=\triangle m_{q} + m_{q}$ where $\triangle m_{q}$ is
small.  In principle, as noted earlier a similar analysis can be
performed for other lattice parameters in addition to the quark mass,
such as the lattice spacing $(a)$, volume $(V)$, the lattice action
etc.

\begin{figure}[!h]
  \begin{center}
 \includegraphics[height=0.46\textwidth,angle=90]{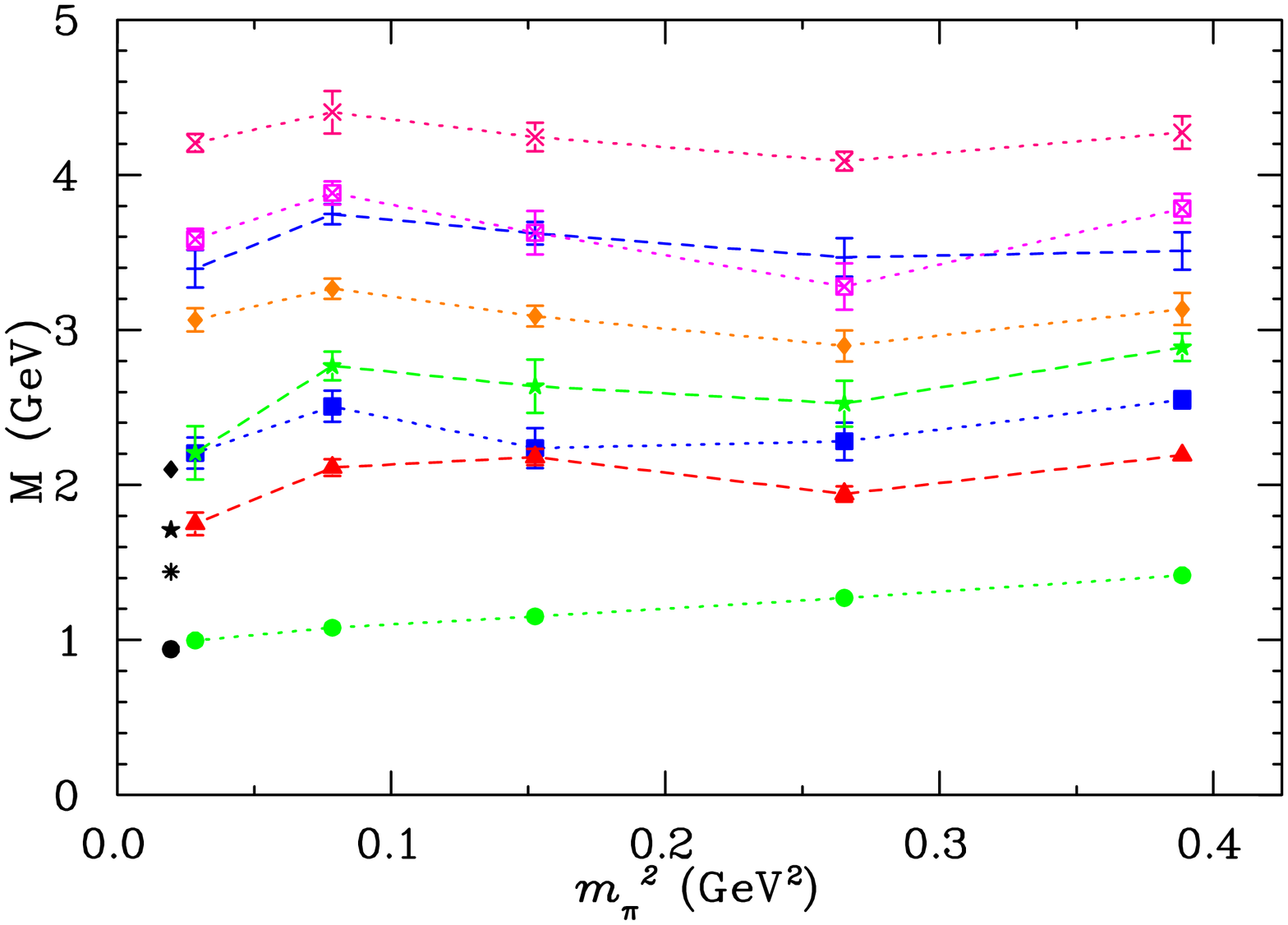}
    \caption{(Color online) As in
      Fig.~\ref{fig:m.x1x2.8x8.forall_kappa}, but with $\chi_{1},\,
      \chi_{4}$ interpolators. The dotted lines in the figure connect
      similar eigenvectors. }
  \label{fig:m.x1x4.8x8.forall_kappa}
  \end{center} 
\end{figure}

The eigenvectors are also tracked for the correlation matrix analysis
with the $\chi_{1}$ and $\chi_{4}$ interpolators and presented in
Fig.~\ref{fig:m.x1x4.8x8.forall_kappa}, which can be compared with
Fig.~\ref{fig:m.x1x2.8x8.forall_kappa}.

\section{Quark-mass Flow of Eigenstates}
 \label{sec:quark_mass_flow}

 \subsection{Positive Parity}
  \label{sub:positive_parity}
A key feature of large correlation matrices is the ability to identify
and isolate energy eigenstates which are nearly degenerate in energy.
However, this approximate degeneracy makes it difficult to trace the
flow of states from one quark mass to the next.  Thus a clear
identification of these near-degenerate states through the features of
the eigenvectors $w^{\alpha}$ isolating the states is necessary in
order to trace the propagation of the states from the heavy to the
light quark-mass region. At this point it is useful to clarify our use
of language. Where we say eigenvector we are referring to the
orthogonal eigenvectors, $w^{\alpha}$, of our symmetric (or hermitian)
$M\times M$ correlation matrix $G(t)$. Where we speak of energy
eigenvalues and energy eigenstates, we are referring to the
eigenvalues and eigenstates of the lattice Hamiltonian, $H$. Our goal
in calculations is always to choose the $M$ interpolators well so that
the few $(<M)$ lowest eigenvalues extracted are a good approximation
to the few lowest energy eigenvalues of $H$ and so that $M$
eigenvectors of our $M\times M$ correlation matrix $G(t)$ capture the
dominant characteristics of the corresponding eigenstate of $H$.

\begin{figure}[!tp]
\begin{center}
\subfloat[Eigenvector components for an $8\times 8$ correlation matrix
  with $\chi_{1},\,\chi_{2}$ interpolators. Odd and even numbers in
  the legend correspond to the $\chi_{1}$ and $\chi_{2}$
  respectively.]  {\label{fig:evectors_8x8_x1x2_sym}
  {\includegraphics[height=0.43\textwidth,angle=90]{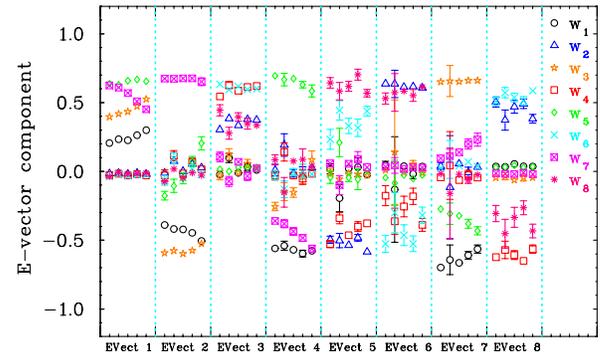}}}\\ 

\subfloat[As in Fig.~\ref{fig:evectors_8x8_x1x2_sym}, but for the
  $\chi_{2}$ and $\chi_{4}$ interpolators. Odd and even numbers in the
  legend correspond to the $\chi_{4}$ and $\chi_{2}$ respectively.]
         {\label{fig:evectors_8x8_x4x2_sym}
           {\includegraphics[height=0.43\textwidth,angle=90]{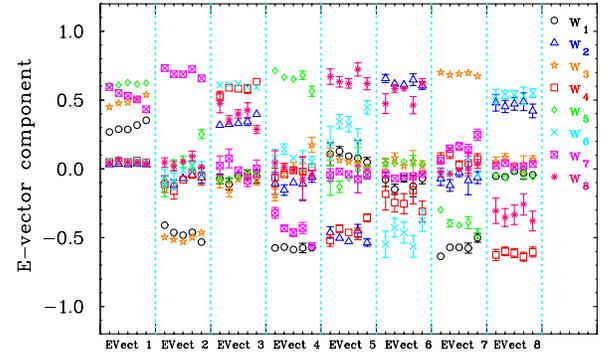}}}\\

\subfloat[As in Fig.~\ref{fig:evectors_8x8_x1x2_sym}, but for the
  $\chi_{1}$ and $\chi_{4}$ interpolators. Odd and even numbers in the
  legend correspond to the $\chi_{1}$ and $\chi_{4}$ respectively.]
         {\label{fig:evectors_8x8_x1x4_sym}
           {\includegraphics[height=0.43\textwidth,angle=90]{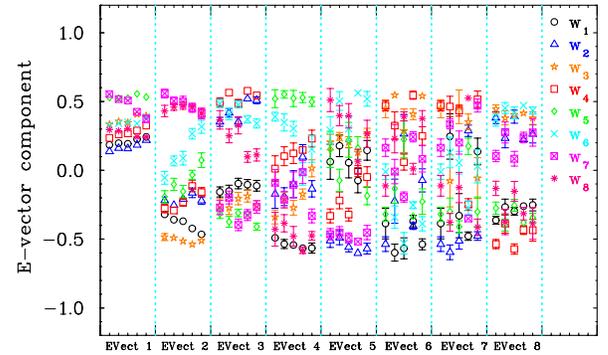}}}\\
 \caption{$\vec{w}^{\alpha}$ is presented for the five different quark
   masses for the $N{\frac{1}{2}}^{+}$ channel after identifying
   eigenvectors via $\vec{w}^{\alpha}(m_{q}) \cdot
   \vec{w}^{\beta}(m_{q^{\prime}})$. For each eigenvector shown in the
   horizontal axis, the eigenvector components are plotted in order of
   increasing quark mass from left to right. Note that Evect 1 to
   Evect 8 correspond to eigenvectors $w^{1}$ to $w^{8}$.  In the
   legend, subscripts $(1,\, 2)$, $(3,\, 4)$, $(5,\,6)$ and $(7,\,8)$
   correspond to the smearing-sweep levels of $16$, $35$, $100$ and
   $200$, respectively.}
 \label{fig:evectors_8x8_x1x2_x4x2_x1x4_sym}
\end{center}
\end{figure}

The anticipated and relatively smooth flow of the eigenvectors as a
function of the quark mass is presented in
Fig.~\ref{fig:evectors_8x8_x1x2_x4x2_x1x4_sym}. It appears that each
eigenvector corresponds to an energy eigenstate of $H$ with the
eigenvector $w^{\alpha}$ capturing some of the core properties of the
corresponding full energy eigenstate of $H$.  While the quark-mass
dependent trends can be significant, our approach reliably allows the
identification of energy eigenstates at adjacent quark masses.

\begin{table*}
 \begin{center}
   \caption{The scalar product $\vec{w}^{\alpha}(m_{q}) \cdot
     \vec{w}^{\beta}(m_{q^\prime})$, for $\kappa = 0.13700$
     ($m_{\pi}=702$ MeV) and $\kappa^\prime = 0.13727$ ($m_{\pi}=572$
     MeV) (top left), $\kappa = 0.13727$ ($m_{\pi}=572$ MeV) and
     $\kappa^\prime = 0.13754$ ($m_{\pi}=402$ MeV) (top right),
     $\kappa = 0.13754$ ($m_{\pi}=402$ MeV) and $\kappa^\prime =
     0.13770$ ($m_{\pi}=293$ MeV) (bottom left), $\kappa = 0.13770$
     ($m_{\pi}=293$ MeV) and $\kappa^\prime = 0.13781$ ($m_{\pi}=156$
     MeV) (bottom right), for an $8\times 8$ correlation matrix of
     $\chi_{1}$ and $\chi_{2}$, with four different levels of
     smearings.  States are ordered from left to right for
     $m_{q^\prime}$ and top to bottom for $m_q$ in order of increasing
     excited-state mass. $\alpha$ and $\beta$ correspond to row and
     column, respectively.}
 \label{table:wdotw_forallkappa_x1x2}
 \vspace{0.3cm}
 \begin{tabular}{cccccccc||cccccccc}
 \hline
 \hline
  \textbf{1.00} & -0.09 & 0.00 & 0.00 & 0.01 & 0.00 & 0.01 & 0.00      &    \textbf{1.00} & -0.08 & 0.01 & -0.01 & 0.01 & 0.01 & 0.00 & 0.00  \\
  0.09 & \textbf{0.99} & -0.07 & 0.13 & -0.01 & 0.00 & 0.01 & 0.00     &    0.08 & \textbf{0.98} & 0.12 & -0.03 & 0.09 & 0.01 & 0.00 & 0.00   \\
  0.01 & 0.07 & \textbf{1.00} & -0.01 & 0.00 & -0.01 & 0.00 & 0.00     &    -0.02 & -0.12 & \textbf{0.99} & -0.08 & 0.00 & 0.00 & 0.00 & -0.01  \\
  -0.01 & -0.13 & 0.02 & \textbf{0.98} & -0.09 & 0.02 & 0.07 & 0.00    &    -0.01 & -0.09 & -0.01 & 0.03 & \textbf{0.99} & -0.10 & 0.00 & 0.00  \\
  0.01 & 0.01 & 0.00 & -0.09 & \textbf{-0.97} & 0.21 & -0.01 & 0.03    &    0.01 & 0.02 & 0.08 & \textbf{0.99} & -0.02 & 0.01 & 0.07 & 0.05   \\
  0.00 & 0.00 & 0.01 & 0.00 & 0.20 & \textbf{0.95} & -0.07 & -0.23     &    0.00 & 0.00 & -0.01 & -0.08 & 0.00 & -0.08 & \textbf{0.99} & 0.07  \\
  -0.01 & 0.00 & 0.00 & -0.07 & 0.01 & 0.07 & \textbf{0.99} & -0.01    &    0.01 & 0.02 & 0.00 & 0.01 & -0.10 & \textbf{-0.99} & -0.08 & 0.03  \\
  0.00 & 0.00 & 0.00 & -0.01 & -0.08 & -0.21 & 0.01 & \textbf{-0.97}   &    0.00 & 0.00 & 0.00 & -0.04 & 0.00 & 0.03 & -0.08 & \textbf{1.00} \\
\hline
  \textbf{1.00} & -0.04 & -0.02 & 0.04 & 0.01 & 0.00 & 0.00 & 0.00       &   \textbf{1.00} & -0.04 & 0.03 & -0.02 & 0.01 & 0.00 & 0.01 & 0.00     \\
  0.03 & \textbf{0.98} & -0.21 & 0.04 & -0.01 & 0.00 & -0.03 & 0.00      &   0.03 & \textbf{0.97} & 0.25 & 0.06 & -0.02 & -0.01 & -0.01 & -0.01   \\
  0.02 & 0.21 & \textbf{0.97} & 0.01 & 0.14 & 0.04 & -0.02 & -0.04       &   -0.03 & -0.24 & \textbf{0.94} & -0.07 & -0.21 & 0.00 & -0.03 & -0.02  \\
  0.01 & -0.01 & -0.13 & -0.37 & \textbf{0.92} & 0.08 & -0.03 & -0.03    &   0.02 & -0.06 & -0.03 & \textbf{0.93} & -0.36 & -0.06 & 0.02 & 0.00   \\
  -0.04 & -0.04 & -0.05 & \textbf{0.93} & 0.36 & 0.02 & 0.00 & -0.01     &   -0.01 & -0.05 & 0.20 & 0.35 & \textbf{0.89} & -0.03 & -0.21 & -0.01  \\
  0.00 & -0.03 & -0.01 & 0.01 & -0.01 & -0.25 & \textbf{-0.97} & -0.03   &   -0.01 & -0.01 & 0.06 & 0.04 & 0.18 & -0.23 & \textbf{0.93} & -0.20   \\
  0.00 & -0.01 & -0.04 & 0.01 & -0.10 & \textbf{0.95} & -0.24 & -0.16    &   0.01 & 0.00 & -0.02 & -0.08 & -0.04 & \textbf{-0.97} & -0.22 & 0.05  \\
  0.00 & -0.01 & -0.02 & 0.00 & -0.02 & -0.15 & 0.07 & \textbf{-0.99}    &   0.01 & 0.00 & -0.04 & -0.02 & -0.04 & 0.00 & -0.20 & \textbf{-0.98}  \\
  \hline
\end{tabular}
\end{center}
\end{table*}

\begin{figure}[!tp]
\begin{center}
\subfloat[Eigenvector components for an $8\times 8$ correlation matrix with
  $\chi_{1},\,\chi_{2}$ interpolators.  Odd and even numbers in the legend correspond to
   the $\chi_{1}$ and $\chi_{2}$ respectively.]
  {\label{fig:evectors_8x8_x1x2}
  {\includegraphics[height=0.43\textwidth,angle=90]{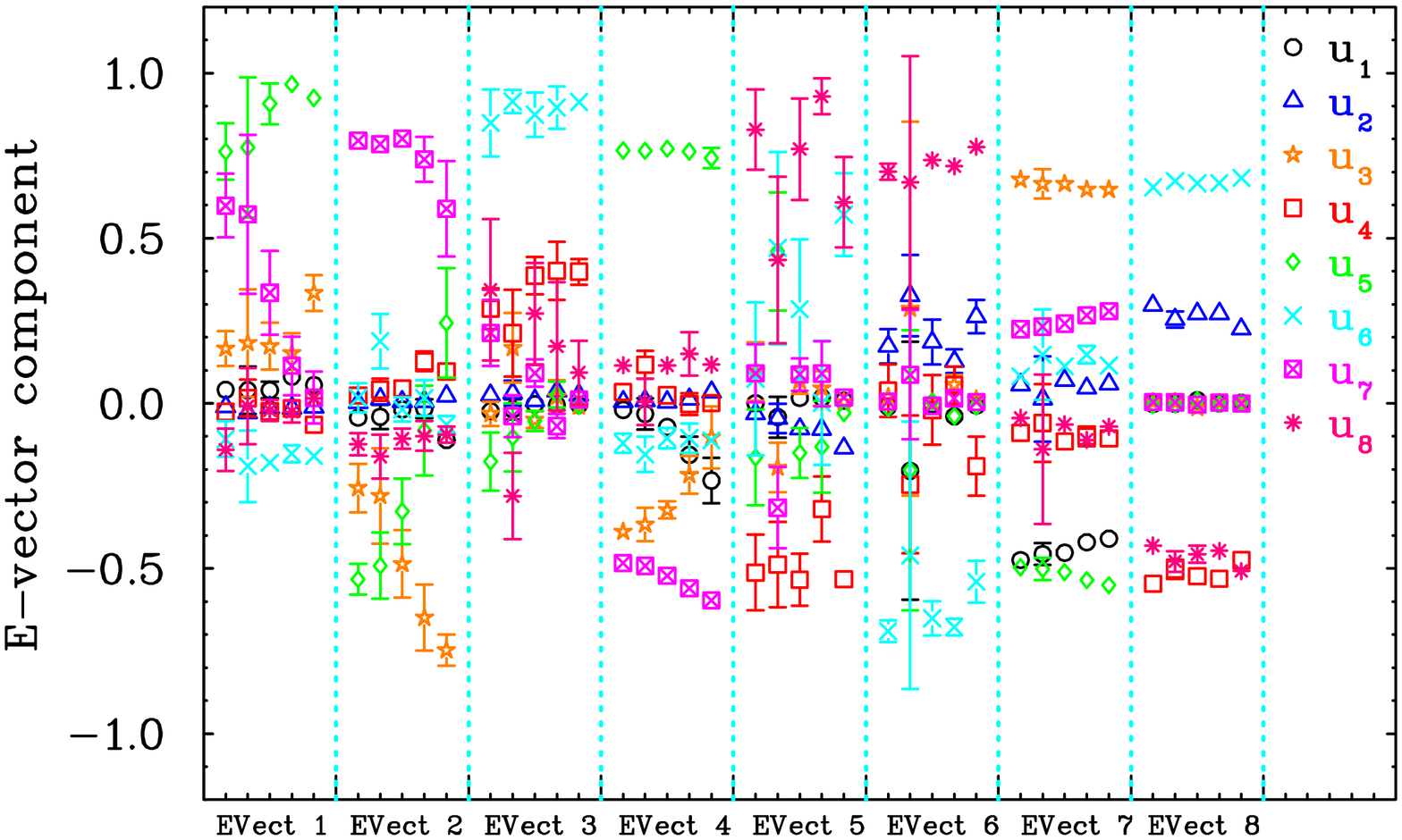}}}\\

\subfloat[As in Fig.~\ref{fig:evectors_8x8_x1x2},
    but for the $\chi_{2}$ and $\chi_{4}$ interpolators. Odd and even numbers
    in the legend correspond to the $\chi_{4}$ and $\chi_{2}$ respectively.]
  {\label{fig:evectors_8x8_x4x2}
  {\includegraphics[height=0.43\textwidth,angle=90]{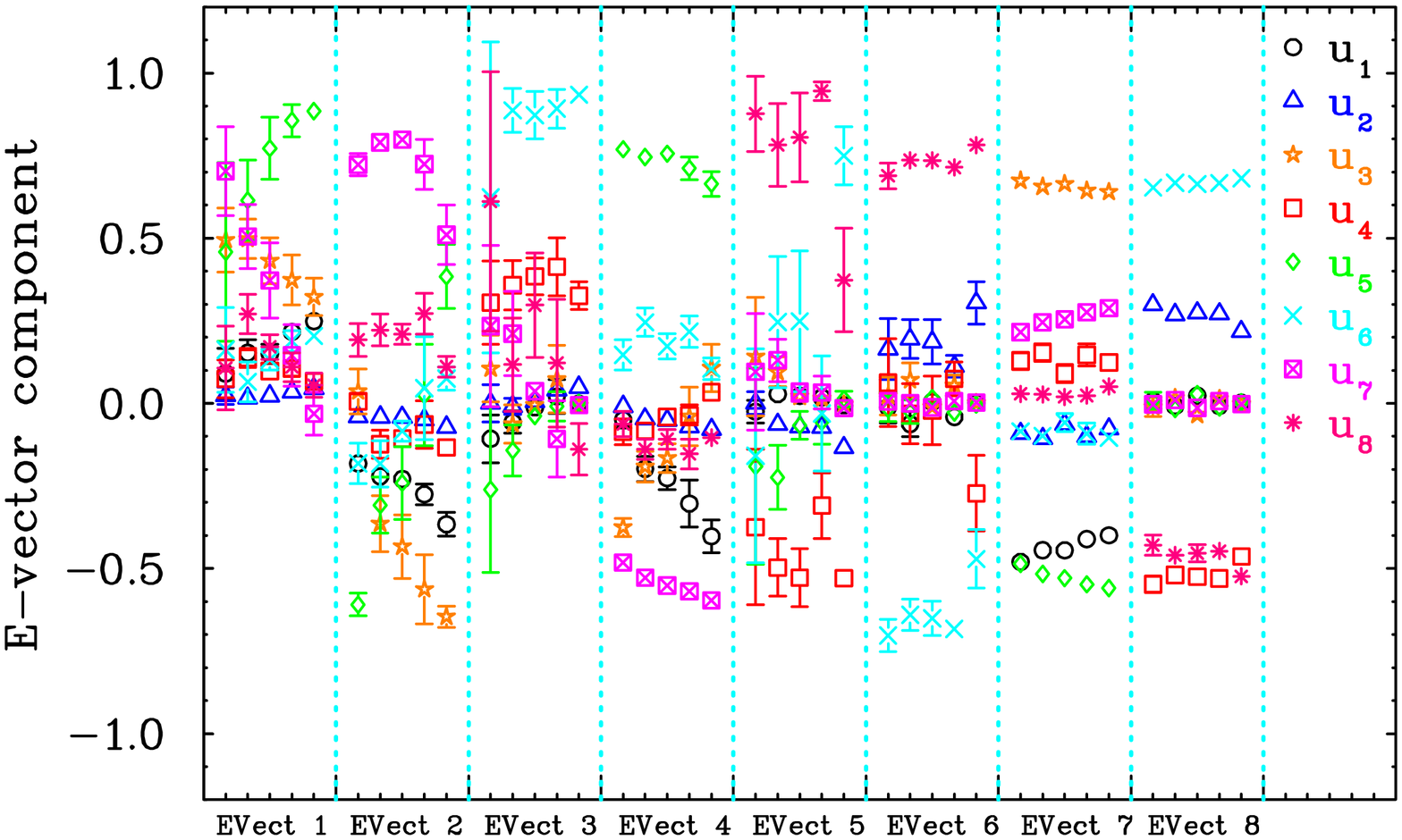}}}\\

\subfloat[As in Fig.~\ref{fig:evectors_8x8_x1x2},
    but for the $\chi_{1}$ and $\chi_{4}$ interpolators. Odd and even numbers in the legend correspond to 
    the $\chi_{1}$ and $\chi_{4}$ respectively.]
  {\label{fig:evectors_8x8_x1x4}
  {\includegraphics[height=0.43\textwidth,angle=90]{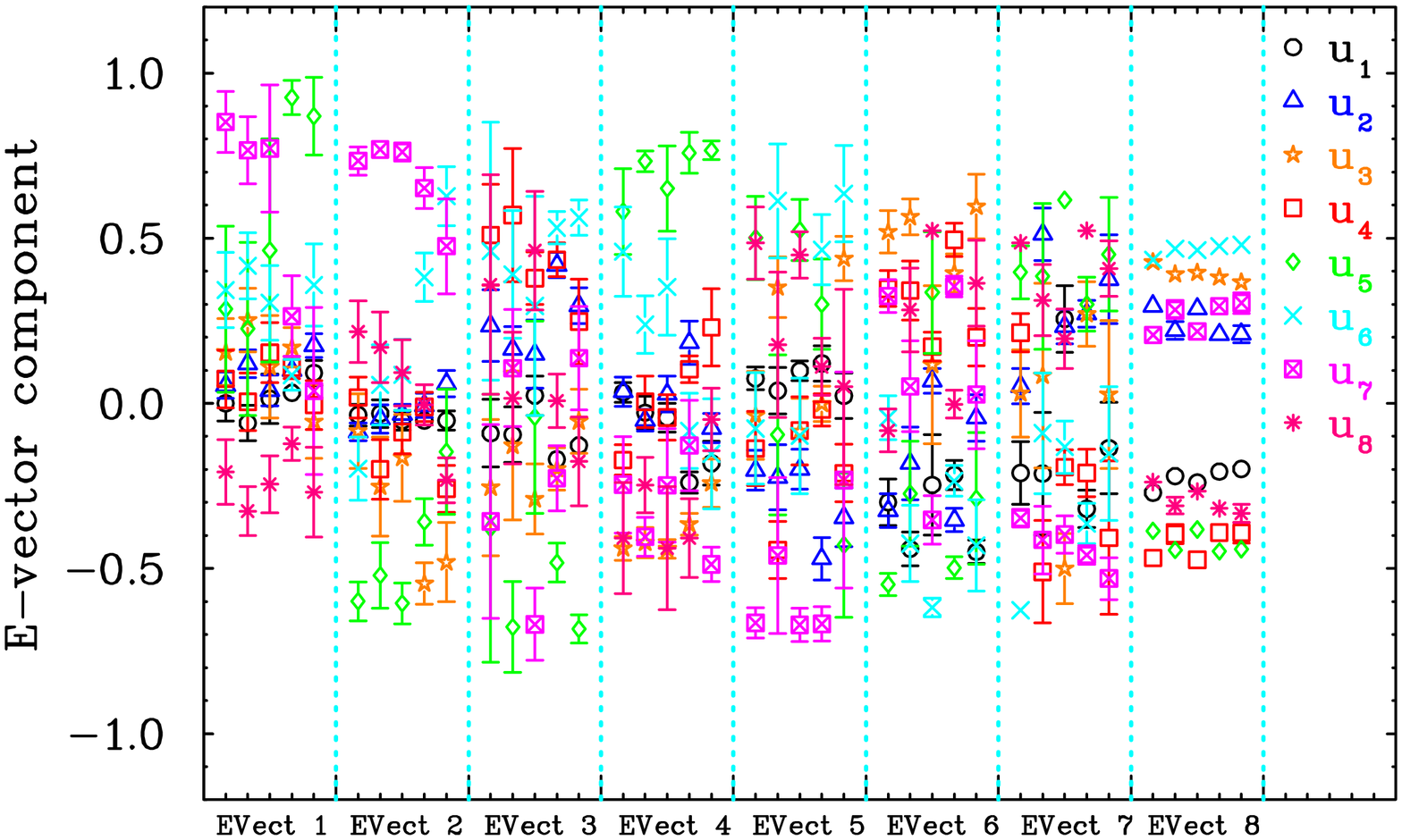}}}

 \caption{$\vec{u}^{\alpha}$ is presented
   for the five different quark masses for the $N{\frac{1}{2}}^{+}$
   channel. For each eigenvector shown in the horizontal axis, the
   eigenvector components are plotted in order of increasing quark
   mass from left to right. Note that Evect 1 to
    Evect 8 correspond to eigenvectors $u^{1}$ to $u^{8}$.
    In the legend, subscripts
  $(1,\, 2)$, $(3,\, 4)$, $(5,\,6)$ and $(7,\,8)$ correspond to the
    smearing-sweep levels of $16$, $35$, $100$ and $200$, respectively.}
 \label{fig:evectors_8x8_x1x2_x2x4_x1x4_asym}
\end{center}
\end{figure}

As the $\chi_{1}$ and $\chi_{4}$ spin-flavor interpolators are very
similar for the $N{\frac{1}{2}}^{+}$ channel, the overall flow of the
eigenvectors $\vec{w}^{\alpha}$ obtained from the $\chi_{1},\chi_{2}$
and $\chi_{2},\chi_{4}$ correlation matrices are very similar in
Fig.~\ref{fig:evectors_8x8_x1x2_x4x2_x1x4_sym}. Also, the overall
strength of the eigenvector components creating and annihilating
$N{\frac{1}{2}}^{+}$ energy-states in the QCD vacuum remains approximately
the same for the $\chi_{1},\chi_{2}$ and $\chi_{2},\chi_{4}$ cases
(Fig.~\ref{fig:evectors_8x8_x1x2_x2x4_x1x4_asym}), which implies that
the eigenstate-energies isolated by the $\chi_{1},\chi_{2}$ and
$\chi_{2},\chi_{4}$ analysis are the same. As the first excited state
is purely $\chi_{1}$-spin-flavour dominated, this state is revealed in
all the three different $8\times 8$ correlation matrix analyses.

There are a few general trends apparent in
Figs.~\ref{fig:evectors_8x8_x1x2_x2x4_x1x4_asym}(a)
and~\ref{fig:evectors_8x8_x1x2_x2x4_x1x4_asym}(b) which are worthy of
note. Focusing on~\ref{fig:evectors_8x8_x1x2_x2x4_x1x4_asym}(a) for
specific reference, we see that there is often competition between
different smearing levels in creating the states. A good example is in
state two, where the 200 sweep $\chi_{1}$ interpolator is complemented
by the 35 sweep $\chi_{1}$ interpolator at heavy quark masses, but the
35 sweep interpolator strength is phased out as one approaches light
quark masses with strength transitioning to the 100 sweep $\chi_{1}$
interpolator.  Even in the ground state one can see the importance of
the 200 sweep interpolator increasing as one approaches the lighter
masses.  This effect is even stronger in
Fig.~\ref{fig:evectors_8x8_x1x2_x2x4_x1x4_asym}(b) for the
$\chi_{4}\chi_{2}$ analysis where the 100 sweep $\chi_{4}$ operator is
phased out in favour of the 200 sweep $\chi_{4}$.

One can also observe the superposition of Gaussian smearings of
different sizes being superimposed with relative minus signs in a
manner that will create nodes in the radial wave function of the
interpolator of Eq.~(\ref{eq:def_bar_phi_alpha}). Focusing again on
Fig.~\ref{fig:evectors_8x8_x1x2_x2x4_x1x4_asym}(a) to provide a specific example,
consider state 2.  Here the widest Gaussian of 200 sweeps is
complemented by a smaller Gaussian with the opposite sign.  Moreover,
as the quark masses become light, the smaller Gaussian grows in size.
The result is that the radial node position of the wave function
increases in distance as the quarks become lighter.

Similarly, state 4 involves the superposition of three Gaussian
smearings with alternating signs.  Here the 200 sweep interpolator is
complemented with the 100 sweep interpolator with opposite sign which
is complemented further by the 35 sweep operator, again with the
opposite sign.  Such a linear combination can create two nodes in the
radial wave function.

Finally, state 7 combines the 200, 100, 35 and 16 sweep interpolators
with alternating signs such that a state with three nodes could be
accessed.

Turning our attention to the $\chi_{1}\chi_{4}$ analysis, we see the
flow of eigenvector components is not as smooth.
Figures~\ref{fig:evectors_8x8_x1x2_x4x2_x1x4_sym}(c)
and~\ref{fig:evectors_8x8_x1x2_x2x4_x1x4_asym}(c) present the flow of the eigenvectors
for this analysis.  A careful comparison of the eigenstate spectrum
with that from the $\chi_{1}\chi_{2}$ analysis at each quark mass and
consideration of the eigenvector flow of the states reveals that the
states dominated by $\chi_{1}$ in the $\chi_{1}\chi_{2}$ analysis are
reproduced in the $\chi_{1}\chi_{4}$ analysis.  However, the remaining
four states display a flow different from those revealed in the
$\chi_{1}\chi_{2}$ analysis.  Thus, a superposition of the
$\chi_{1},\chi_{2}$ and $\chi_{1},\chi_{4}$ analysis provides $12$
unique energy states.

\begin{figure}
  \begin{center}
 \includegraphics[height=0.48\textwidth,angle=90]{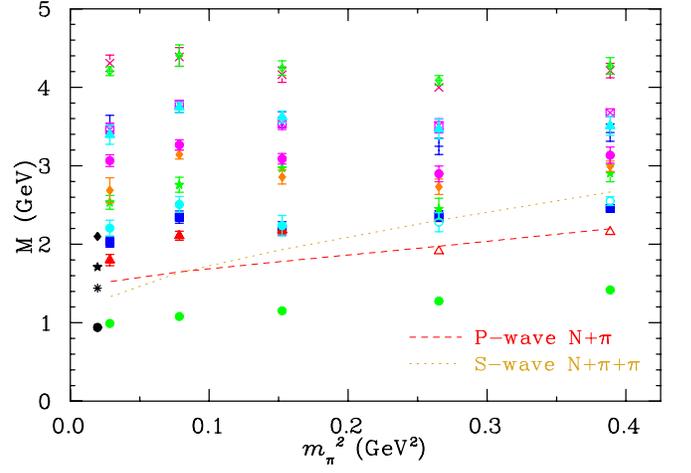}
    \caption{(Color online) $N{\frac{1}{2}}^{+}$ energy-states from
      $8\times 8\times 2$ correlation matrices of
      $\chi_{1},\chi_{2},\chi_{4}$~\cite{Mahbub:2010rm}. The $p$-wave
      $N \pi$ scattering threshold (with one unit of lattice momentum)
      and the $s$-wave $N\pi\pi$ threshold are presented by dashed and
      dotted lines, respectively.}
\label{fig:m.8x8x2.pswave}
  \end{center} 
\end{figure}

In Fig.~\ref{fig:m.8x8x2.pswave}, a superposition of the two $8\times8$
 analysis $(8\times 8 \times 2)$ of $\chi_{1},\chi_{2}$ and
$\chi_{1},\chi_{4}$ is presented.  Scattering $p$-wave $\pi N$ and
$s$-wave $\pi\pi N$ energy levels are also shown.

For the two large quark masses, as seen in
Fig.~\ref{fig:m.8x8x2.pswave}, the extracted lattice results sit close
to the scattering two particle $p$-wave $N\pi$ threshold
($E_{N}+E_{\pi}$) with back-to-back momenta, $\vec p = (2\pi/ L_x, 0,
0)$ and $s$-wave $N\pi\pi$ threshold ($M_{N}+M_{\pi}+M_{\pi}$),
whereas the masses for the lighter three quarks sit much higher. There
is no evidence of these scattering states at light quark masses. Our
conclusion is that our 3-quark operators have very little coupling to
the multi-hadron states relative to the states we do observe at the
light quark masses.

It is noted that the couplings to the multi-particle meson-baryon
states are suppressed by $1/\sqrt{V}$ relative to states dominated by
a single-particle state. Due to the large volume of our lattice, it is
likely that multi-particle states will be suppressed and missed in our
spectrum, particularly at lighter quark masses where the quark-mass
effect also acts to suppress the spectral strength.  Further analysis
of finite volume effects~\cite{Young:2002cj} on the spectrum is highly
desirable. Future calculations should also investigate the use of
five-quark operators to ensure better overlap with the multi-particle
states and to better resolve and probe the excited state
spectrum~\cite{Morningstar:2011ka,Lang:2012db,Morningstar:2013bda}.

\subsection{Negative Parity}
 \label{sub:negative_parity}

Now we can repeat our analysis for the
identification of the $N{\frac{1}{2}}^{-}$ states.  In
Table~\ref{table:wdotw_forallkappa_x1x2_negP}, the scalar product
${\cal{W}}^{\alpha\beta}=\vec{w}^{\alpha}(m_{q}) \cdot
\vec{w}^{\beta}(m_{q^\prime})$ for all the quark masses is presented.

\begin{figure}
  \begin{center}
 \includegraphics[height=0.46\textwidth,angle=90]{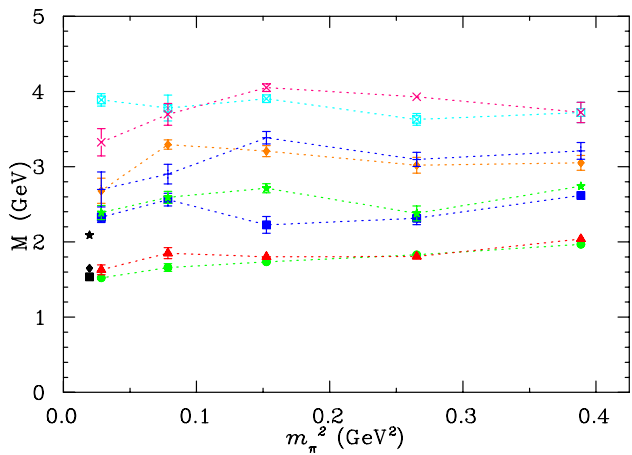}
    \caption{(Color online) Masses of $N{\frac{1}{2}}^{-}$
      energy-states from an $8\times 8$ correlation matrix of
      $\chi_{1}, \chi_{2}$ in
      Table~\ref{table:wdotw_forallkappa_x1x2_negP}. The
      dotted lines in the figure illustrates 
      the eigenvector flow.}
\label{fig:m.x1x2.8x8.negP}
  \end{center} 
\end{figure}

\begin{figure}
  \begin{center}
 \includegraphics[height=0.46\textwidth,angle=90]{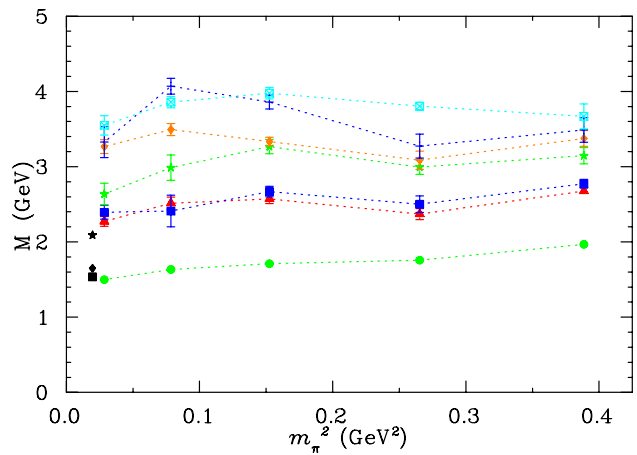}
    \caption{(Color online) As in Fig.~\ref{fig:m.x1x2.8x8.negP}, but
      with $\chi_{1}, \chi_{4}$ interpolators. The dotted lines in the
      figure illustrates the eigenvector flow. }
\label{fig:m.x1x4.8x8.negP}
  \end{center} 
\end{figure}

In Figs.~\ref{fig:m.x1x2.8x8.negP} and \ref{fig:m.x1x4.8x8.negP},
the $N{\frac{1}{2}}^{-}$ spectrum from the $8\times 8$ analysis involving
$\chi_{1} ,\chi_{2}$ and $\chi_{1},\chi_{4}$ is presented,
respectively. While the $\chi_{1},\chi_{4}$ analysis is able to
extract a low-lying energy state, it misses the near-degenerate second
energy state in this channel. This second energy-state is revealed in
the $\chi_{1},\,\chi_{2}$ spin-flavor combination presenting
two nearly-degenerate low-lying states, which is in accord with the
quark model based on $SU(6)$ spin-flavor symmetry. Recall that three
spin-$\frac{1}{2}$ quarks may provide a total spin of
$s=\frac{1}{2}$ or $\frac{3}{2}$, the $L=1$ state can couple two
different ways to provide a $J=\frac{1}{2}$ state, hence providing two
orthogonal spin-$\frac{1}{2}$ states in the $L=1$, 70 plet
representation of $SU(6)$. Both of these states have a width of
$\approx$150 MeV.

\begin{figure}[!tp]
\begin{center}
\subfloat[Eigenvector components for an $8\times 8$ correlation matrix with
  $\chi_{1},\,\chi_{2}$ interpolators. Odd and even numbers in the legend  correspond to 
    the $\chi_{1}$ and $\chi_{2}$ respectively.]
  {\label{fig:evectors_8x8_x1x2_sym_negP}
  {\includegraphics[height=0.43\textwidth,angle=90]{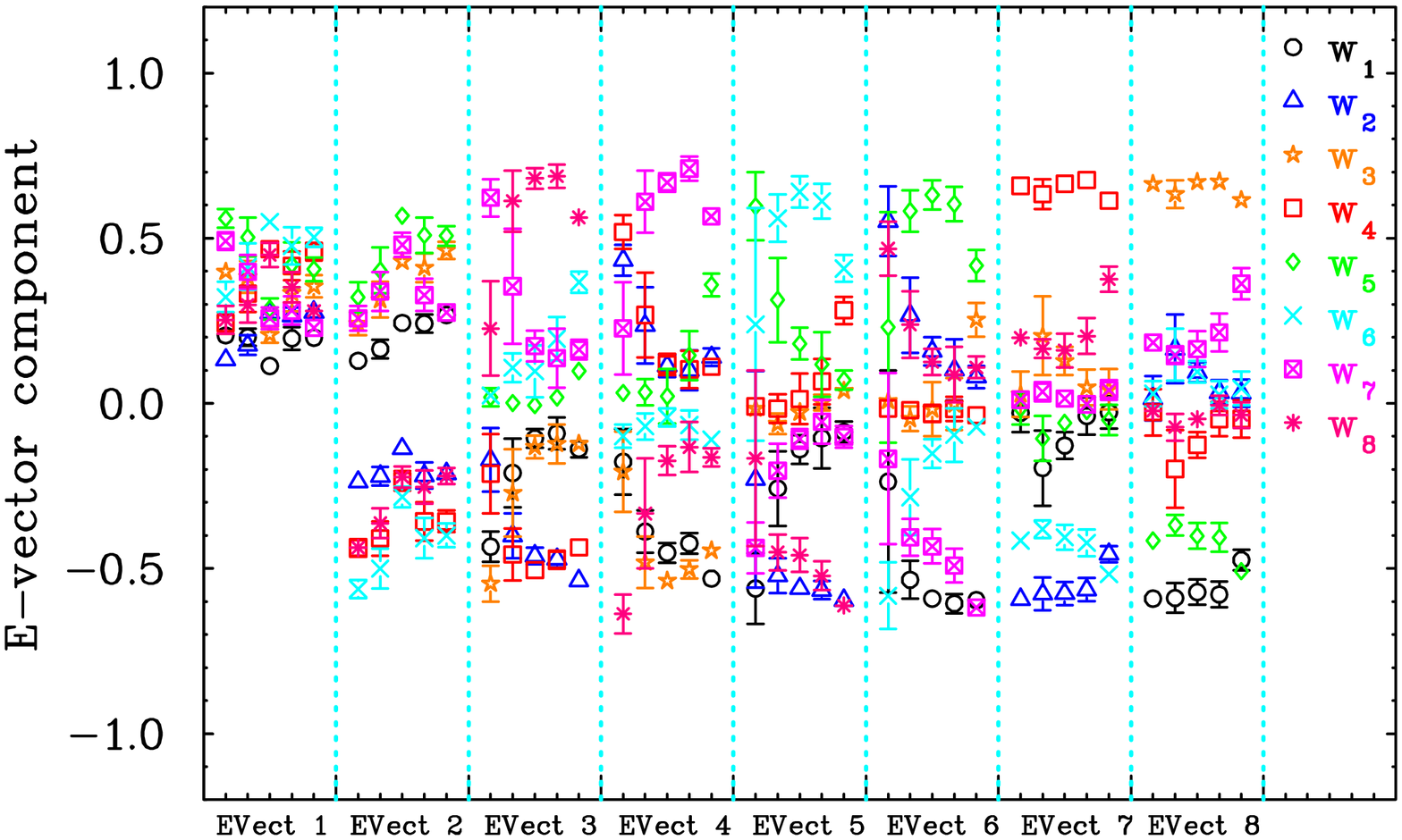}}}\\ %&

\subfloat[As in Fig.~\ref{fig:evectors_8x8_x1x2_sym_negP},
    but for $\chi_{2}$ and $\chi_{4}$ interpolators. Odd and even numbers in the legend  correspond to 
    the $\chi_{4}$ and $\chi_{2}$ respectively.]
  {\label{fig:evectors_8x8_x4x2_sym_negP}
  {\includegraphics[height=0.43\textwidth,angle=90]{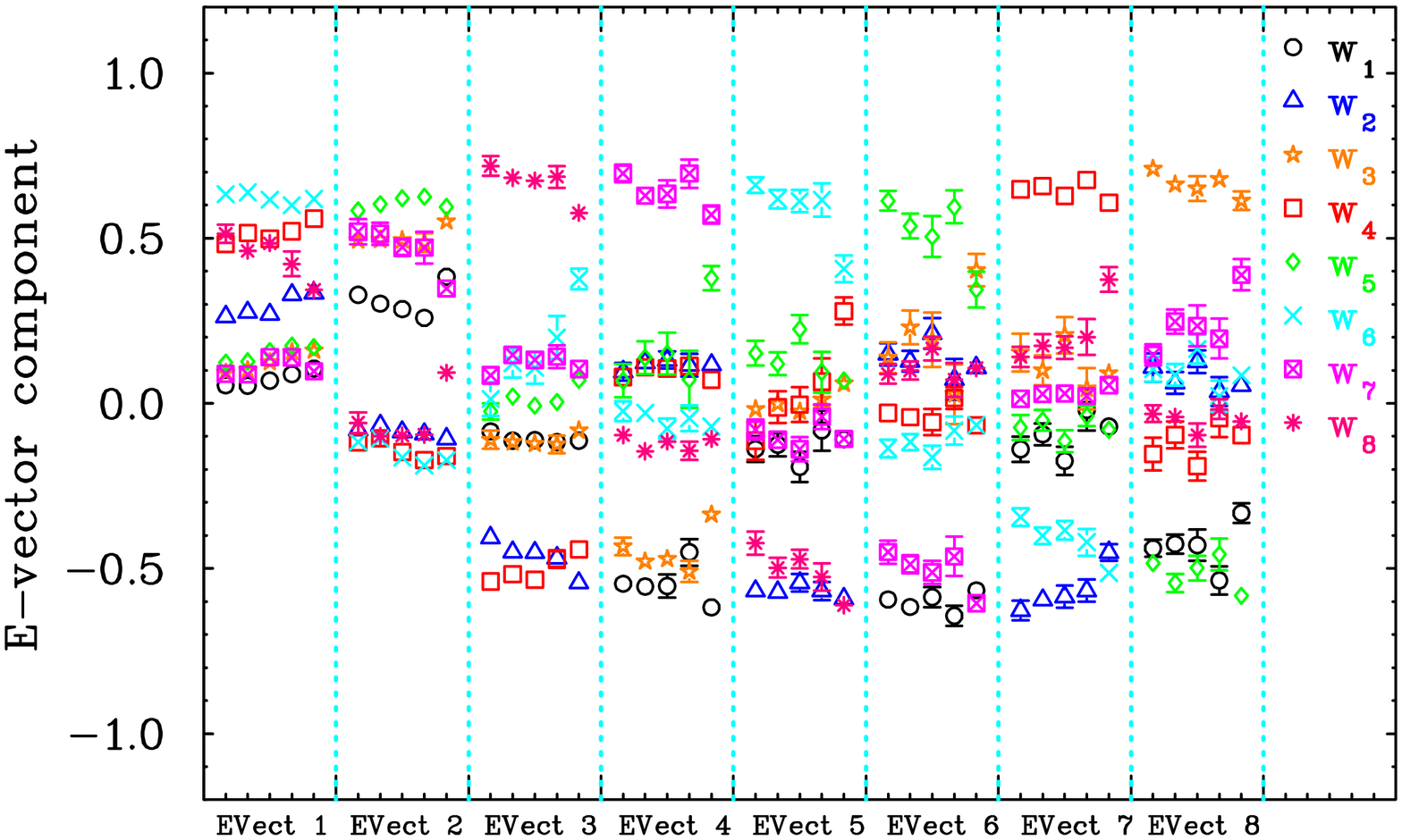}}}\\

\subfloat[As in Fig.~\ref{fig:evectors_8x8_x1x2_sym_negP},
    but for $\chi_{1}$ and $\chi_{4}$ interpolators. Odd and even numbers in the legend  correspond to 
    the $\chi_{1}$ and $\chi_{4}$ respectively.]
  {\label{fig:evectors_8x8_x1x4_sym_negP}
  {\includegraphics[height=0.43\textwidth,angle=90]{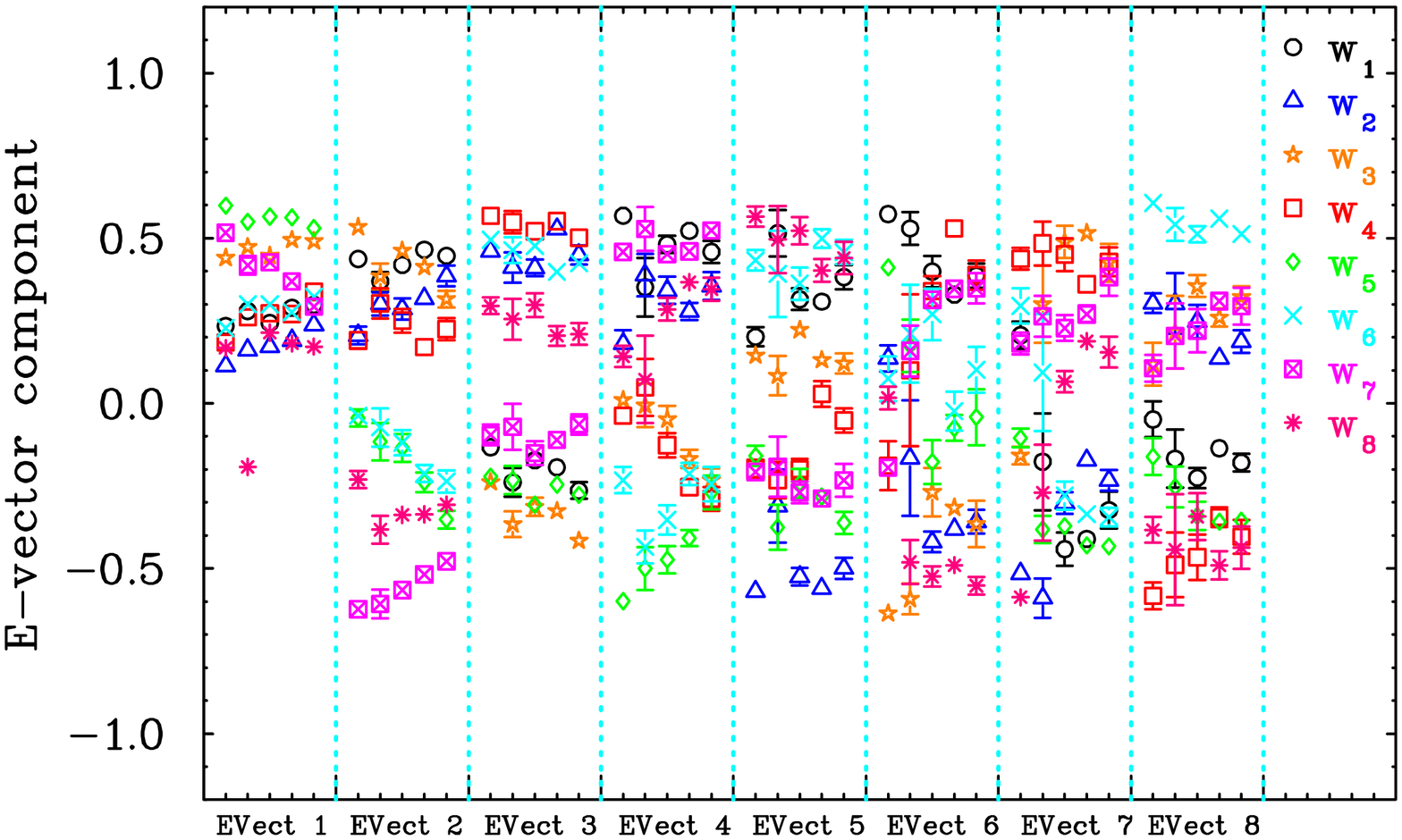}}}\\
 \caption{$\vec{w}^{\alpha}$ for the five different quark masses are
   presented for the $N{\frac{1}{2}}^{-}$ channel after identifying
   eigenvectors via $\vec{w}^{\alpha}(m_{q}) \cdot
   \vec{w}^{\beta}(m_{q^{\prime}})$. For each eigenvector shown in
   horizontal axis, the eigenvector components are plotted in order of
   increasing quark mass from left to right. Note that Evect 1 to
   Evect 8 correspond to eigenvectos $w^{1}$ to $w^{8}$.
   In the legend, subscripts $(1,\, 2)$, $(3,\,
   4)$, $(5,\,6)$ and $(7,\,8)$ correspond to the smearing-sweep
   levels of $16$, $35$, $100$ and $200$ respectively.}
 \label{fig:evectors_8x8_x1x2_x4x2_x1x4_sym_negP}
\end{center}
\end{figure}

\begin{figure}[!tp]
\begin{center}
\subfloat[Eigenvector components for an $8\times 8$ correlation matrix with
  $\chi_{1},\,\chi_{2}$ interpolators. Odd and even numbers in the legend correspond to
   the $\chi_{1}$ and $\chi_{2}$ respectively.]
  {\label{fig:evectors_8x8_x1x2_asym_negP}
  {\includegraphics[height=0.43\textwidth,angle=90]{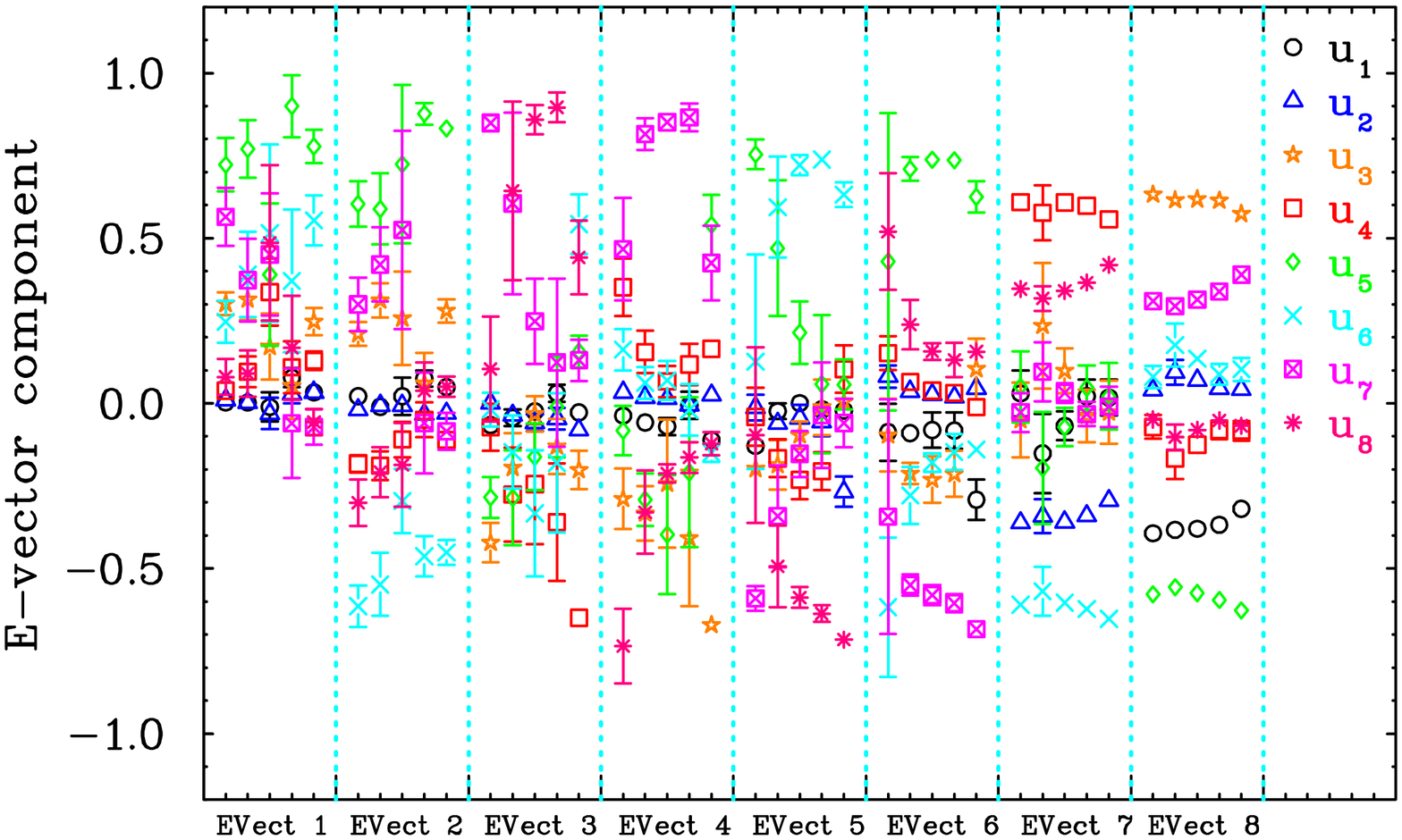}}}\\ %&

\subfloat[As in Fig.~\ref{fig:evectors_8x8_x1x2_asym_negP},
    but for $\chi_{2}$ and $\chi_{4}$ interpolators. Odd and even numbers in the legend correspond to
   the $\chi_{4}$ and $\chi_{2}$ respectively. ]
  {\label{fig:evectors_8x8_x4x2_asym_negP}
  {\includegraphics[height=0.43\textwidth,angle=90]{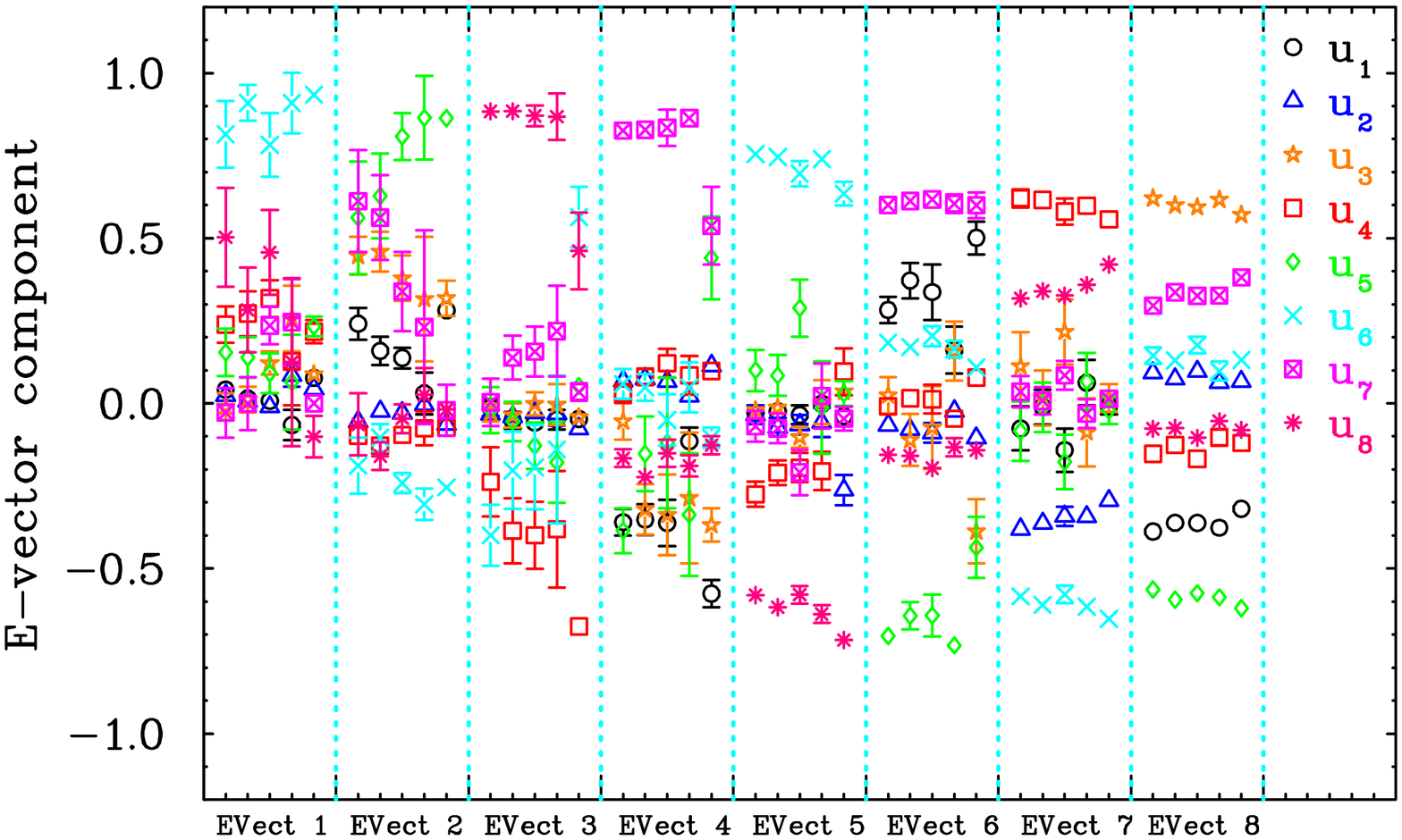}}}\\

\subfloat[As in Fig.~\ref{fig:evectors_8x8_x1x2_asym_negP},
    but for $\chi_{1}$ and $\chi_{4}$ interpolators. Odd and even numbers in the legend correspond to
   the $\chi_{1}$ and $\chi_{4}$ respectively. ]
  {\label{fig:evectors_8x8_x1x4_asym_negP}
  {\includegraphics[height=0.43\textwidth,angle=90]{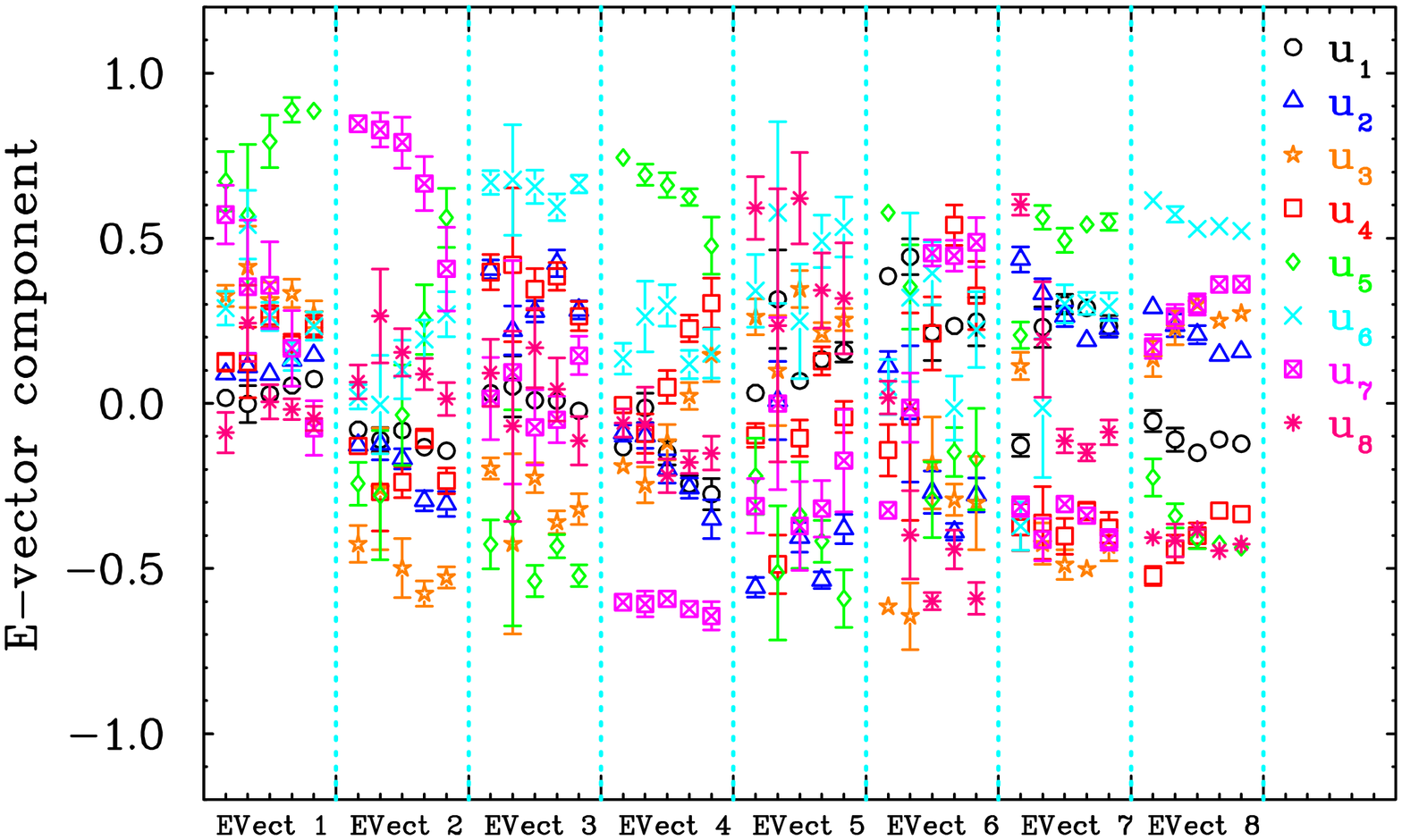}}}\\
 \caption{$\vec{u}^{\alpha}$ for the five different quark masses are
   presented for the $N{\frac{1}{2}}^{-}$ channel. For each
   eigenvector shown in horizontal axis, the eigenvector components
   are plotted in order of increasing quark mass from left to
   right. Note that Evect 1 to Evect 8 correspond to eigenvectos
   $u^{1}$ to $u^{8}$. In the legend,
   subscripts $(1,\, 2)$, $(3,\, 4)$, $(5,\,6)$ and $(7,\,8)$
   correspond to the smearing-sweep levels of $16$, $35$, $100$ and
   $200$ respectively.}
 \label{fig:evectors_8x8_x1x2_x4x2_x1x4_asym_negP}
\end{center}
\end{figure}

As in Fig.~\ref{fig:evectors_8x8_x1x2_x4x2_x1x4_sym}, the eigenvector
components for different quark masses are presented in
Figs.~\ref{fig:evectors_8x8_x1x2_x4x2_x1x4_sym_negP} and
\ref{fig:evectors_8x8_x1x2_x4x2_x1x4_asym_negP} for the
$N{\frac{1}{2}}^{-}$ channel. It is interesting to note that the
scalar-diquark interpolator $\chi_{1}$ dominates the lowest two $N^{-}$
states (Fig.~\ref{fig:evectors_8x8_x1x2_x4x2_x1x4_asym_negP}) when available.
The $\chi_{2}$ interpolator makes an important contribution in
creating the second energy state in this channel. 

\begin{figure}
  \begin{center}
 \includegraphics[height=0.46\textwidth,angle=90]{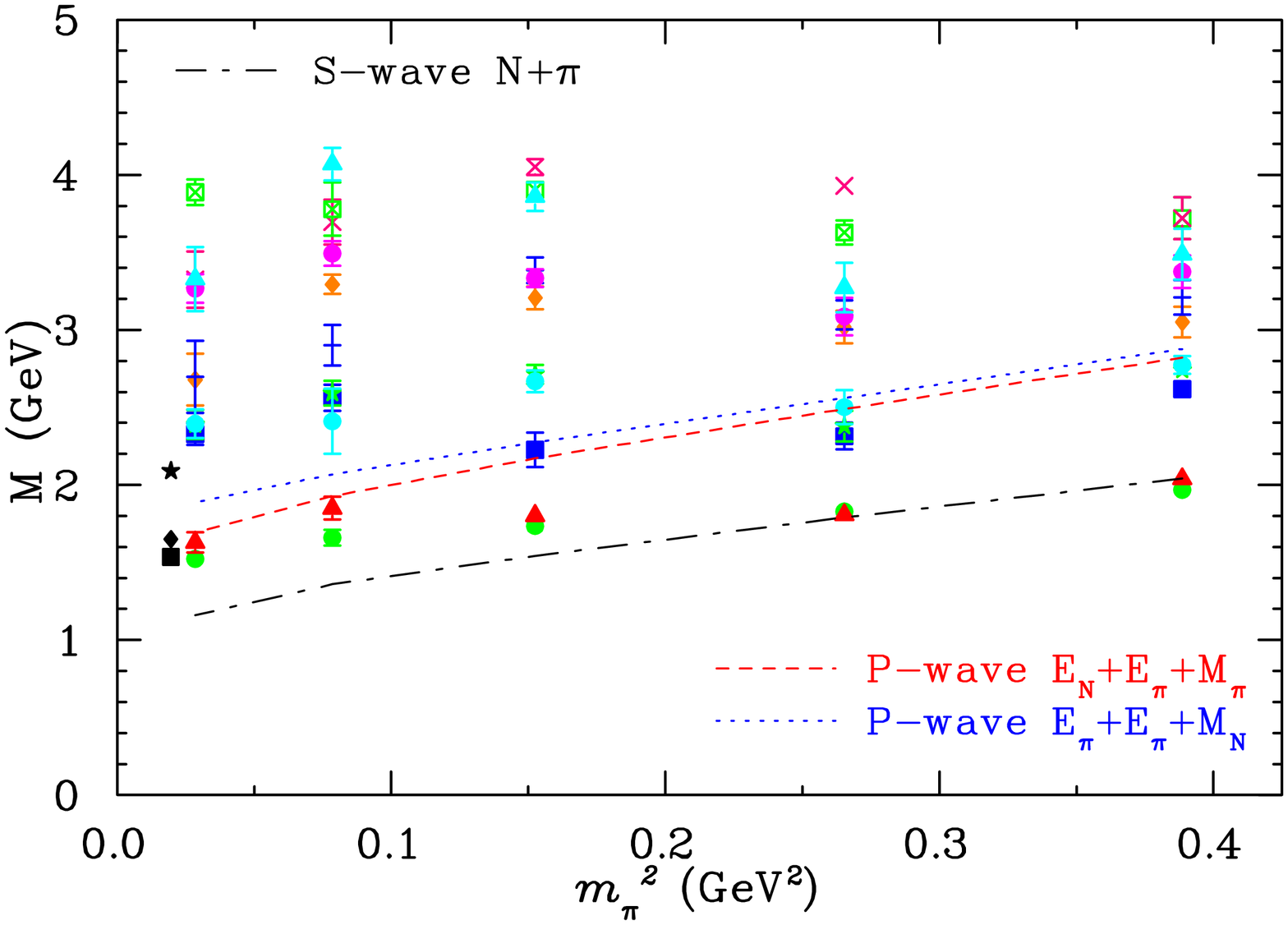} 
    \caption{(Color online) Masses of $N{\frac{1}{2}}^{-}$
      energy-states from $8\times 8 \times 2$ correlation matrices of $\chi_{1},
      \chi_{2}, \chi_{4} $.}
\label{fig:m-.12x12}
  \end{center} 
\end{figure}

In Fig.~\ref{fig:m-.12x12}, a superposition of the two $8\times 8$
analysis $(8\times 8 \times 2)$ of $\chi_{1},\chi_{2}$ and
$\chi_{1},\chi_{4}$ is presented.  Scattering $p$-wave $\pi N$ and
$s$-wave $\pi\pi N$ states are also shown.  The results for the lowest
energy-state at the two heavier pion masses sit close to the
scattering s-wave $N+\pi$ ($M_{\pi}+M_{N}$) threshold indicating that
these results may be scattering states at these pion masses. However,
they disappear from our spectrum at the light pion masses. A similar
situation also prevails the second energy-state, where the state sits
close to the $p$-wave $E_{N}+E_{\pi}+M_{\pi}$ and
$E_{\pi}+E_{\pi}+M_{N}$ scattering threshold with back-to-back momenta
of one lattice unit, $\vec p = (2 \pi/ L_x, 0, 0)$. Again the use of
5- or 7- etc. quark meson-baryon operators will be required to explore
these scattering
states~\cite{Morningstar:2011ka,Lang:2012db,Morningstar:2013bda}.

\begin{table*}
 \begin{center}
 \caption{The scalar product $\vec{w}^{\alpha}(m_{q}) \cdot
   \vec{w}^{\beta}(m_{q^\prime})$, for $\kappa = 0.13700$
   ($m_{\pi}=702$ MeV) and $\kappa^\prime = 0.13727$ ($m_{\pi}=572$
   MeV) (top left), $\kappa = 0.13727$ ($m_{\pi}=572$ MeV) and
   $\kappa^\prime = 0.13754$ ($m_{\pi}=402$ MeV) (top right), $\kappa
   = 0.13754$ ($m_{\pi}=402$ MeV) and $\kappa^\prime = 0.13770$
   ($m_{\pi}=293$ MeV) (bottom left), $\kappa = 0.13770$
   ($m_{\pi}=293$ MeV) and $\kappa^\prime = 0.13781$ ($m_{\pi}=156$
   MeV) (bottom right), for an $8\times 8$ correlation matrix of
   $\chi_{1}$ and $\chi_{2}$, with four different levels of smearings,
   for the $N{\frac{1}{2}}^{-}$ states.  States are ordered from left
   to right for $m_{q^\prime}$ and top to bottom for $m_q$ in order of
   increasing excited-state mass. $\alpha$ and $\beta$ correspond to
   row and column, respectively.}
 \label{table:wdotw_forallkappa_x1x2_negP}
 \vspace{0.3cm}
 \begin{tabular}{cccccccc||cccccccc}
 \hline
 \hline
  -0.03 & \textbf{0.99} & -0.08 & -0.04 & 0.05 & -0.01 & 0.00 & 0.01       &    -0.22 & \textbf{0.97} & 0.04 & -0.10 & -0.01 & 0.04 & 0.01 & -0.02     \\
  \textbf{1.00} & 0.02 & 0.01 & -0.08 & -0.02 & -0.01 & 0.00 & 0.01        &    \textbf{0.97} & 0.22 & -0.07 & -0.07 & 0.01 & 0.02 & -0.02 & -0.01     \\
  0.00 & 0.07 & \textbf{0.97} & 0.05 & 0.22 & -0.03 & -0.04 & 0.00         &    0.07 & -0.03 & \textbf{0.99} & -0.04 & 0.07 & 0.00 & -0.02 & -0.01     \\
  0.07 & 0.03 & -0.07 & \textbf{0.96} & 0.01 & -0.27 & 0.00 & -0.02        &    0.05 & 0.11 & 0.04 & \textbf{0.99} & -0.01 & 0.05 & 0.01 & -0.04       \\
  -0.02 & 0.06 & 0.20 & 0.02 & \textbf{-0.95} & -0.01 & -0.23 & -0.01      &    -0.01 & 0.02 & -0.07 & 0.02 & \textbf{0.99} & -0.09 & 0.04 & 0.03      \\
  0.03 & 0.02 & 0.01 & 0.26 & 0.00 & \textbf{0.94} & -0.01 & -0.21         &    0.02 & 0.04 & 0.01 & 0.04 & -0.09 & \textbf{-0.99} & -0.02 & -0.02     \\
  -0.01 & 0.02 & 0.08 & 0.01 & -0.22 & 0.01 & \textbf{0.97} & 0.01         &    0.02 & -0.01 & 0.01 & -0.01 & -0.04 & -0.01 & \textbf{0.99} & -0.13    \\
  0.00 & 0.00 & 0.00 & 0.08 & 0.00 & 0.20 & -0.02 & \textbf{0.98}          &    0.01 & 0.02 & 0.02 & 0.03 & -0.03 & -0.02 & 0.12 & \textbf{0.99}       \\
\hline
    \textbf{0.91} & 0.40 & 0.02 & 0.02 & 0.01 & -0.05 & 0.00 & 0.00       &   \textbf{0.98} & 0.17 & -0.06 & -0.01 & 0.01 & 0.01 & 0.00 & 0.00      \\
    0.40 & \textbf{-0.91} & 0.00 & 0.01 & -0.02 & 0.01 & -0.01 & 0.00     &   -0.17 & \textbf{0.99} & -0.01 & -0.03 & 0.01 & 0.01 & -0.01 & 0.00    \\
    -0.01 & -0.01 & \textbf{0.96} & -0.27 & 0.01 & -0.01 & 0.00 & 0.02    &   0.05 & 0.03 & \textbf{0.77} & 0.64 & -0.01 & 0.00 & 0.00 & 0.02       \\
    -0.03 & 0.00 & 0.27 & \textbf{0.96} & 0.01 & 0.01 & 0.02 & 0.00       &   0.04 & -0.01 & 0.64 & \textbf{-0.77} & -0.04 & 0.00 & 0.01 & 0.00     \\
    0.04 & 0.03 & 0.01 & -0.01 & -0.22 & \textbf{0.97} & 0.02 & 0.01      &   0.00 & 0.00 & 0.02 & -0.01 & 0.66 & \textbf{-0.75} & -0.04 & -0.01    \\
    0.01 & -0.01 & -0.01 & -0.01 & \textbf{0.98} & 0.22 & 0.04 & 0.00     &   -0.01 & -0.01 & 0.03 & -0.02 & \textbf{0.75} & 0.66 & -0.04 & -0.02   \\
    0.00 & 0.00 & -0.02 & 0.01 & 0.01 & -0.01 & -0.12 & \textbf{0.99}     &   0.00 & 0.01 & 0.00 & 0.01 & 0.05 & -0.01 & \textbf{0.96} & -0.26      \\
    0.01 & -0.01 & 0.00 & -0.02 & -0.04 & -0.03 & \textbf{0.99} & 0.12    &   0.00 & 0.00 & -0.01 & -0.01 & 0.04 & 0.01 & 0.26 & \textbf{0.96}      \\
  \hline
\end{tabular}
\end{center}
\end{table*}

\section{Conclusions}
\label{section:conclusions}

In this paper, a comprehensive analysis for the nucleon spectrum with
$I=1/2$, $s=1/2$, is presented using the correlation matrix
approach. In particular, a method for energy-eigenstate identification
and flow is presented and demonstrated for the positive and negative
parity channel. Details of the method developed for an identification
and the propagation of the energy-states from heavy to light quark
mass region are provided. In particular, the new technique is useful
in identifying the flow of the near-degenerate energy eigenstates from
one quark mass to the next.  The eigenvectors obtained from the
eigenvalue equations for several $8\times 8$ correlation matrices are
utilized in tracking the eigen-energy states.

In presenting the results, both non-symmetric left and right
eigenvalue equations and a symmetric eigenvalue equation are
considered. While the masses are the same in the two different
approaches, the eigenvectors obtained from the symmetric matrix are
orthogonal.  Thus the generalized measure ${\mathcal W}^{\alpha\beta}$
is used to track the flow of eigenvectors with quark mass.  The scalar
product of the eigenvectors shows its robustness in tracking the flow
of the energy eigenstates even when the energies are nearly
degenerate.

The coefficients of the interpolators creating and annihilating a
state in the QCD vacuum are also presented. The flow of the
eigenvectors reveals a smooth pattern and presents important insights
into baryon structure and its evolution with quark mass.

Another interesting result of this paper is that, the correlation
matrix method can be used to track the energy-states that are involved
in an avoided level crossing. It is noted that the avoided level
crossings lie within the error bars, but the demonstration of the
robustness of the approach remains.

Future steps include the introduction of five-quark meson-baryon
operators in the correlation matrices, to ensure the clear isolation
of states and ultimately extract the resonance parameters from the
first principles of QCD.

\begin{acknowledgments}
We thank PACS-CS Collaboration for making these $2+1$ flavor
configurations available. This research was undertaken on the NCI
National Facility in Canberra, Australia, which is supported by the
Australian Commonwealth Government. We also acknowledge eResearch SA
for generous grants of supercomputing time which have enabled this
project.  This research is supported by the Australian Research
Council.
\end{acknowledgments}

\appendix*

\section{Pedagogical Discussion of the Correlation Matrix}

\subsection{For $M=N$}
Let us consider an $N$ dimensional Hilbert space with a hamiltonian, $\hat{H}$, and let
$\vert \chi_{1}\rangle$, $\vert \chi_{2}\rangle$, $\cdots$, $\vert
\chi_{N}\rangle$ be $N$ linearly independent states. Similarly, let
$\vert E_{1}\rangle$, $\vert E_{2}\rangle$, $\cdots$, $\vert
E_{N}\rangle$ be a complete orthonormal basis of energy eigenstates, then the state
$\vert\chi_{i}\rangle$ can be written as
\begin{align}
 \vert\chi_{i}\rangle &=
 \sum_{j=1}^{N}\vert{E_{j}}\rangle\langle{E_{j}}\vert\chi_{i}\rangle = \sum_{j=1}^{N}C_{ij}\vert{E_{j}}\rangle,
 \label{eq:statepsi}
\end{align}
where, $C_{ij}\equiv \langle{E_{j}}\vert\chi_{i}\rangle$, which is
analogous to Eq.~(\ref{eq:def_lambdabar_posp}). In matrix form
Eq.~(\ref{eq:statepsi}) can be written as
\begin{align}
  \left[\begin{array}{c}
  \vert\chi_{1}\rangle \\
  \vdots \\
  \vert\chi_{N}\rangle
 \end{array}\right]
=[C]\left[\begin{array}{c}
 \vert{E_{1}}\rangle \\
 \vdots \\
 \vert{E_{N}}\rangle 
 \end{array}\right]
\end{align}
Since $\vert\chi_{i}\rangle$ is linearly independent, then $C$ must be non-singular and so there must exist
$(C^{-1})_{ij}$ such that 
\begin{align}
  \left[\begin{array}{c}
  \vert E_{1}\rangle \\
  \vdots \\
  \vert E_{N}\rangle
 \end{array}\right]
=[C^{-1}]\left[\begin{array}{c}
 \vert\chi_{1}\rangle \\
 \vdots \\
 \vert\chi_{N}\rangle 
 \end{array}\right].
\end{align}
Similarly, $\langle\chi_{i}\vert$ can be expressed as
$\langle\chi_{i}\vert=\sum_{j=1}^{N}\langle\chi_{i}\vert
E_{j}\rangle\langle E_{j}\vert=\sum_{j=1}^{N}C_{ij}^{\ast}\langle
E_{j}\vert$. 

Let us now define
$\vert\chi_{i}\rangle\equiv\hat\chi_{i}^{\dagger}\vert\Omega\rangle$
for $N$ linearly independent field operators $(\hat\chi_{i}^{\dagger})$, where the
dagger denotes the adjoint. Then we may write
\begin{align}
G_{ij}(t)&\equiv \langle\Omega\vert\hat\chi_{i}(t)\hat\chi_{j}^{\dagger}(0)\vert\Omega\rangle
\nonumber \\
 & = \langle\Omega\vert\hat\chi_{i}
 e^{-i\hat{H}t}\hat\chi_{j}^{\dagger}\vert\Omega\rangle \nonumber \\
 & = \langle\chi_{i}\vert e^{-i\hat{H}t}\vert\chi_{j}\rangle \nonumber \\
 &= \sum_{k,l=1}^{N}C_{ik}^{\ast}\langle E_{k}\vert
 e^{-i\hat{H}t}\vert E_{l}\rangle C_{jl} \nonumber \\
 &= \sum_{k=1}^{N}C_{ik}^{\ast}e^{-E_{k}t}C_{jk}  \nonumber \\
 &= \sum_{k=1}^{N}C_{ik}^{\ast}e^{-E_{k}t}(C^{\ast\dagger})_{kj}  \nonumber 
\end{align}
or,
\begin{align}
 [G(t)] &= [C^{\ast}]\left(\begin{array}{ccc} e^{-E_{1}t} & 0
     & 0  \\ 0 & e^{-E_{2}t} & 0 \\ 0 & 0 &
     \ddots \end{array}\right)[C^{\ast\dagger}] \nonumber \\
       &\equiv  [C^{\ast}][E(t)][C^{\ast\dagger}],
\label{eq:cstar_Et_cstardagg}
\end{align} 
where, $[E(t)]$ is obviously diagonal. The
Eq.~(\ref{eq:cstar_Et_cstardagg}) is analogous to
Eq.~(\ref{eq:cor_at_zero_mom}). From this point on for notational convenience
we will no longer use square brackets to denote matrices. We see that,
$G(t)=G(t)^{\dagger}$ for all $t$.  From
Eq.~(\ref{eq:cstar_Et_cstardagg}), we can also write
\begin{align}
G(t_{0}) &= C^{\ast}\sqrt{E(t_{0})}\sqrt{E(t_{0})}C^{\ast\dagger} \nonumber \\
           &= C^{\prime\ast} C^{\prime\ast\dagger},
\end{align}
where, $C^{\prime\ast}\equiv C^{\ast}\sqrt{E(t_{0})}$ and
$C^{\prime\ast\dagger} \equiv
\sqrt{E(t_{0})}C^{\ast\dagger}$. Similarly, we can write
\begin{align}
G(t_{0}+\triangle t) &= C^{\prime\ast}E(\triangle t) C^{\prime\ast\dagger}.
 \label{eq:cpstar_Edeltat_cpstardagg}
\end{align}

\noindent
Let us consider the polar decomposition of $C^{\prime\ast}$ as \\
$C^{\prime\ast}=\sqrt{C^{\prime\ast}C^{\prime\ast \dagger}}\,U=\sqrt{G(t_{0})}\,U$, where $U$ is unitary
\begin{align}
 UU^{\dagger}= U^{\dagger}U = I. \nonumber 
\end{align}
Similarly, $C^{\prime\ast\dagger}=U^{\dagger}\,\sqrt{G(t_{0})}$. Then, Eq.~(\ref{eq:cpstar_Edeltat_cpstardagg}) can be written as
\begin{align}
G(t_{0}+\triangle t) &= \sqrt{G(t_{0})}\,U E(\triangle t) U^{\dagger}\,\sqrt{G(t_{0})}
\end{align}
or equivalently
\begin{align}
\tilde{G}(\triangle t) & \equiv \sqrt{G(t_{0})}^{-1}G(t_{0}+\triangle t)\sqrt{G(t_{0})}^{-1} \nonumber \\ 
                       &= U E(\triangle t) U^{\dagger},
 \label{eq:sqrtgt0_gt0plusdeltat_sqrtgt0}
\end{align}
where we are using the notation
\begin{align}
E(\triangle t) = \left[\begin{array}{ccc} e^{-E_{1}\triangle t} & 0
     & 0  \\ 0 & e^{-E_{2}\triangle t} & 0 \\ 0 & 0 &
     \ddots \end{array}\right].
\end{align}
Denoting the normalized eigenvectors
of $\tilde{G}(\triangle t)$ as $\vec{w}^{i}$ for
$i=1,\cdots,N$, then $U$ consists of columns $U=[\vec{w}^{1}\vert
  \vec{w}^{2}\vert \cdots \vert\vec{w}^{N}]$.
Hence we see that
\begin{align}
\tilde{G}(\triangle t)\vec{w}^{i}=e^{-E_{i}\triangle t}\vec{w}^{i}.
 \label{eq:evalue_eq_with_gtilde}
\end{align}
Multiplying Eq.~(\ref{eq:evalue_eq_with_gtilde}) by $G(t_{0})^{-1/2}$ from left and defining 
\begin{align}\vec{u}^{i}\equiv G(t_{0})^{-1/2}\vec{w}^{i},\end{align}
gives
\begin{align}
 G(t_{0})^{-1}G(t_{0}+\triangle t)\vec{u}^{i}= e^{-E_{i}\triangle t}\vec{u}^{i},
\end{align}
which is analogous to Eq.~(\ref{eqn:right_eigenvalue_equation}).

\subsection{For $M<N$}
Let $\hat\chi_{1}$, $\hat\chi_{2}$, $\cdots$, $\hat\chi_{N}$ be $M$
linearly independent interpolating field operators with $M<N$. Then we
may consider these $M$ $\hat\chi_{i}$'s as a subset of a complete set
of $N$ interpolating operators.  Then as before, we can define
$G^{(M)}(t)$ as an $M\times M$ correlation matrix, then $G^{M}(t)$ can
be written as the upper left $M\times M$ block of $G(t)$ such that
\begin{align}
 G(t)&=\left[\begin{array}{cc}
    G^{M}(t) & G^{m}(t)  \\
    {G^{m}(t)^{\dagger}} & {G^{(N-M)}(t)} \end{array}
\right],
\end{align}
where the off-diagonal rectangular matrix $G^{m}(t)$ has elements
$G^{m}_{ij}(t)$ for $i=1,\cdots,M$ and $j=M+1,\cdots,N$. Clearly
$G^{m}(t)$ mixes the upper and lower diagonal blocks.

In terms of the full $N\times N$ correlation matrix $G(t)$ we define
$\tilde{G}(\triangle t)$ as before and it has the form
\begin{align}
 \tilde{G}(\triangle t)
=\left[\begin{array}{cc}
    {\tilde{G}^{M}(\triangle t)} & \tilde{G}_{m}(\triangle t)  \\
    \tilde{G}_{m}(\triangle t)^{\dagger} & {\tilde{G}^{(N-M)}(\triangle t)}
\end{array}\right], \nonumber
\end{align}
which is of course diagonalized by the full $N\times N$ unitary matrix
$U$ as shown in Eq.~(\ref{eq:sqrtgt0_gt0plusdeltat_sqrtgt0}).

Let us now temporarily assume that the off-diagonal elements of $\tilde{G}(\triangle t)$ are zero, then
\begin{align}
 \tilde{G}(\triangle t)
=\left[\begin{array}{cc}
    {\tilde{G}^{M}(\triangle t)} & {0}  \\
    {0} & {\tilde{G}^{(N-M)}(\triangle t)}
\end{array}\right], \nonumber
\end{align}
and then it follows that the time independent unitary matrix can be written as
\begin{align}
 U=\left[\begin{array}{cc}
    {U^{M}} & {0} \\
    {0}     & {U^{(N-M)}}
\end{array}\right]. \nonumber
\end{align}
This will occur if and only if we have chosen our $M$ interpolating
fields such that they span exactly the same subspace as $M$ of the
exact energy eigenstates.

But we will certainly never exactly achieve this and so mixing will
occur through $G^{m}(t)\ne 0$. In this case $U$ will not have this
convenient block diagonal form.  Let $U^{\prime}$ diagonalize
$\tilde{G}^{M}(\triangle t)$, i.e. $U^{\prime}$ is an $M\times M$
unitary matrix defined at this particular $\triangle t$. Then if
$G^{m}(t)\ne 0$ we see that $U^{\prime}\ne U^{M}$ and in general
$U^{\prime}$ will not be independent of $\triangle t$. Note that $M$
linearly independent interpolating field operators will operate on the
vacuum and give rise to $M$ linearly independent states. The more
closely these $M$ linearly independent states come to spanning the
same subspace as $M$ exact energy eigenstates, then the smaller will
be $G^{m}(t)$ and the more block diagonal will be $\tilde{G}(\triangle
t)$ and $U$. We can only ensure that $\tilde{G}^{M}(\triangle t)$
commutes with itself at all $\triangle t$ if $U$ is block diagonal,
i.e., if $G^{m}(t)=0$ so that $U^{\prime}=U^{M}$ and is therefore time
independent. Then the smaller will be the off-diagonal elements and
the smaller will be the mixing and the excited state contamination.

In practice $M<<N$, since on a lattice the dimensionality of Hilbert
space is in the many millions. We would like to choose the $M$
interpolating fields such that they span an $M$-dimensional subspace
that has the greatest overlap possible with the subspace spanned by
the $M$ lowest energy eigenstates $\vert E_{1}\rangle,\, \cdots,
\vert E_{M}\rangle$. With the lowest energy eigenstates our numerical errors will
be minimized, since by working at large Euclidean times we minimize
the influence of the higher excited state contamination and so
optimize our extraction of the lowest $M$ energy eigenstates.

Since the interpolating fields produced states $\vert \chi_{i}\rangle$
that were not normalized in any way, it is numerically convenient to
redefine
\begin{align}
\hat\chi_{i}(t)\rightarrow
\hat\chi_{i}^{\prime}(t)\equiv
\frac{1}{\sqrt{{G}^{M}_{ii}(0)}}\hat\chi_{i}(t), \nonumber
\end{align}
\noindent
where there is no sum over $i$. Therefore, we attempt to put the
strengths of our interpolating fields at a comparable level by
defining
 \begin{align}
{G}^{\prime M}_{ij}(t)=
\frac{1}{\sqrt{{G}^{M}_{ii}(0)}}
  {G}^{M}_{ij}(t)\frac{1}{\sqrt{{G}^{M}_{jj}(0)}},  \nonumber
\end{align}
to ensure that the matrix elements of ${G}^{\prime M}(t)$ are all
$\sim{\cal{O}}(1)$ to maximize numerical significance of all ``$ij$''
combinations. This is completely legitimate as it is simply a change
to the normalization chosen for our interpolating fields, which we are
free to do. The matrix ${G}^{\prime M}(t)$ is hermitian except for the
effects of finite ensemble and round-off errors.
\noindent
Let us define
\begin{align}
{\hat{G}}^{M}(t)=\frac{1}{2}({G}^{\prime M}(t)+{G}^{\prime M\dagger}(t)),
\end{align}
as an improved unbiased estimator of the ensemble average for $G^{\prime M}(t)$. Then
${\hat{G}}^{M}(t)$ is exactly hermitian. Therefore, as before, we may
diagonalize
\begin{align}
\tilde{\hat{G}}^{M}(\triangle t) \equiv [{\hat{G}^{M}(t_{0})}^{-1/2}{\hat{G}}^{M}(t_{0}+\triangle
  t){\hat{G}^{M}(t_{0})}^{-1/2}],
  \label{eq:final_hermitian_mat}
\end{align}  
which is also hermitian, as ${\hat{G}^{M}(t_{0})}^{-1/2}$ and
${\hat{G}}^{M}(t_{0}+\triangle t)$ are obviously hermitian. The
eigenvectors ($\vec{w}^{\alpha}$) obtained from diagonalizing the
above matrix are therefore orthonormal. It is noted that using the
$U+U^{\ast}$ trick~\cite{Melnitchouk:2002eg}, where these $U$'s are
links here, the hermitian correlation matrix $G^{M}(t)$ is real
symmetric and so the eigenvalues remain real and the eigenvectors
orthogonal.  Again, it is important to note that the more poorly we
choose our $M$ interpolating fields then the bigger will be the
off-diagonal elements of ${\tilde{\hat{G}}}(t)$. Hence the more
time-dependent will be the $U^{\prime}$ and the less reliable our
extracted energies and eigenvectors.

Now we can consider $\hat{G}^{ M}(t)$ as a function of quark mass, $m_{q}$,
to identify how a given state evolves with $m_{q}$.  If
$m_{q}^{\prime}=m_{q}+\triangle m_{q}$ with $\triangle m_{q}$ small,
we expect $\vec{w}^{i}(m_{q}).\vec{w}^{j}(m_{q}^{\prime})\sim
1$ if $i = j$ and
$\vec{w}^{i}(m_{q}).\vec{w}^{j}(m_{q}^{\prime}) \sim 0$ if
$i\ne j$.


\begin{thebibliography}{51}%
\makeatletter
\providecommand \@ifxundefined [1]{%
 \@ifx{#1\undefined}
}%
\providecommand \@ifnum [1]{%
 \ifnum #1\expandafter \@firstoftwo
 \else \expandafter \@secondoftwo
 \fi
}%
\providecommand \@ifx [1]{%
 \ifx #1\expandafter \@firstoftwo
 \else \expandafter \@secondoftwo
 \fi
}%
\providecommand \natexlab [1]{#1}%
\providecommand \enquote  [1]{``#1''}%
\providecommand \bibnamefont  [1]{#1}%
\providecommand \bibfnamefont [1]{#1}%
\providecommand \citenamefont [1]{#1}%
\providecommand \href@noop [0]{\@secondoftwo}%
\providecommand \href [0]{\begingroup \@sanitize@url \@href}%
\providecommand \@href[1]{\@@startlink{#1}\@@href}%
\providecommand \@@href[1]{\endgroup#1\@@endlink}%
\providecommand \@sanitize@url [0]{\catcode `\\12\catcode `\$12\catcode
  `\&12\catcode `\#12\catcode `\^12\catcode `\_12\catcode `\%12\relax}%
\providecommand \@@startlink[1]{}%
\providecommand \@@endlink[0]{}%
\providecommand \url  [0]{\begingroup\@sanitize@url \@url }%
\providecommand \@url [1]{\endgroup\@href {#1}{\urlprefix }}%
\providecommand \urlprefix  [0]{URL }%
\providecommand \Eprint [0]{\href }%
\providecommand \doibase [0]{http://dx.doi.org/}%
\providecommand \selectlanguage [0]{\@gobble}%
\providecommand \bibinfo  [0]{\@secondoftwo}%
\providecommand \bibfield  [0]{\@secondoftwo}%
\providecommand \translation [1]{[#1]}%
\providecommand \BibitemOpen [0]{}%
\providecommand \bibitemStop [0]{}%
\providecommand \bibitemNoStop [0]{.\EOS\space}%
\providecommand \EOS [0]{\spacefactor3000\relax}%
\providecommand \BibitemShut  [1]{\csname bibitem#1\endcsname}%
\let\auto@bib@innerbib\@empty
%</preamble>
\bibitem [{\citenamefont {Durr}\ \emph {et~al.}(2008)\citenamefont {Durr} \emph
  {et~al.}}]{Durr:2008zz}%
  \BibitemOpen
  \bibfield  {author} {\bibinfo {author} {\bibfnamefont {S.}~\bibnamefont
  {Durr}} \emph {et~al.},\ }\href {\doibase 10.1126/science.1163233} {\bibfield
   {journal} {\bibinfo  {journal} {Science}\ }\textbf {\bibinfo {volume}
  {322}},\ \bibinfo {pages} {1224} (\bibinfo {year} {2008})},\ \Eprint
  {http://arxiv.org/abs/0906.3599} {arXiv:0906.3599 [hep-lat]} \BibitemShut
  {NoStop}%
%%CITATION = 0906.3599;%%
\bibitem [{\citenamefont {Roper}(1964)}]{Roper:1964zz}%
  \BibitemOpen
  \bibfield  {author} {\bibinfo {author} {\bibfnamefont {L.~D.}\ \bibnamefont
  {Roper}},\ }\href {\doibase 10.1103/PhysRevLett.12.340} {\bibfield  {journal}
  {\bibinfo  {journal} {Phys. Rev. Lett.}\ }\textbf {\bibinfo {volume} {12}},\
  \bibinfo {pages} {340} (\bibinfo {year} {1964})}\BibitemShut {NoStop}%
%%CITATION = PRLTA,12,340;%%
\bibitem [{\citenamefont {Isgur}\ and\ \citenamefont
  {Karl}(1977)}]{Isgur:1977ef}%
  \BibitemOpen
  \bibfield  {author} {\bibinfo {author} {\bibfnamefont {N.}~\bibnamefont
  {Isgur}}\ and\ \bibinfo {author} {\bibfnamefont {G.}~\bibnamefont {Karl}},\
  }\href {\doibase 10.1016/0370-2693(77)90074-0} {\bibfield  {journal}
  {\bibinfo  {journal} {Phys. Lett.}\ }\textbf {\bibinfo {volume} {B72}},\
  \bibinfo {pages} {109} (\bibinfo {year} {1977})}\BibitemShut {NoStop}%
%%CITATION = PHLTA,B72,109;%%
\bibitem [{\citenamefont {Isgur}\ and\ \citenamefont
  {Karl}(1979)}]{Isgur:1978wd}%
  \BibitemOpen
  \bibfield  {author} {\bibinfo {author} {\bibfnamefont {N.}~\bibnamefont
  {Isgur}}\ and\ \bibinfo {author} {\bibfnamefont {G.}~\bibnamefont {Karl}},\
  }\href {\doibase 10.1103/PhysRevD.19.2653} {\bibfield  {journal} {\bibinfo
  {journal} {Phys. Rev.}\ }\textbf {\bibinfo {volume} {D19}},\ \bibinfo {pages}
  {2653} (\bibinfo {year} {1979})}\BibitemShut {NoStop}%
%%CITATION = PHRVA,D19,2653;%%
\bibitem [{\citenamefont {Li}\ \emph {et~al.}(1992)\citenamefont {Li},
  \citenamefont {Burkert},\ and\ \citenamefont {Li}}]{Li:1991yba}%
  \BibitemOpen
  \bibfield  {author} {\bibinfo {author} {\bibfnamefont {Z.-p.}\ \bibnamefont
  {Li}}, \bibinfo {author} {\bibfnamefont {V.}~\bibnamefont {Burkert}}, \ and\
  \bibinfo {author} {\bibfnamefont {Z.-j.}\ \bibnamefont {Li}},\ }\href
  {\doibase 10.1103/PhysRevD.46.70} {\bibfield  {journal} {\bibinfo  {journal}
  {Phys. Rev.}\ }\textbf {\bibinfo {volume} {D46}},\ \bibinfo {pages} {70}
  (\bibinfo {year} {1992})}\BibitemShut {NoStop}%
%%CITATION = PHRVA,D46,70;%%
\bibitem [{\citenamefont {Carlson}\ and\ \citenamefont
  {Mukhopadhyay}(1991)}]{Carlson:1991tg}%
  \BibitemOpen
  \bibfield  {author} {\bibinfo {author} {\bibfnamefont {C.~E.}\ \bibnamefont
  {Carlson}}\ and\ \bibinfo {author} {\bibfnamefont {N.~C.}\ \bibnamefont
  {Mukhopadhyay}},\ }\href {\doibase 10.1103/PhysRevLett.67.3745} {\bibfield
  {journal} {\bibinfo  {journal} {Phys. Rev. Lett.}\ }\textbf {\bibinfo
  {volume} {67}},\ \bibinfo {pages} {3745} (\bibinfo {year}
  {1991})}\BibitemShut {NoStop}%
%%CITATION = PRLTA,67,3745;%%
\bibitem [{\citenamefont {Guichon}(1985)}]{Guichon:1985ny}%
  \BibitemOpen
  \bibfield  {author} {\bibinfo {author} {\bibfnamefont {P.~A.~M.}\
  \bibnamefont {Guichon}},\ }\href {\doibase 10.1016/0370-2693(85)90341-7}
  {\bibfield  {journal} {\bibinfo  {journal} {Phys. Lett.}\ }\textbf {\bibinfo
  {volume} {B164}},\ \bibinfo {pages} {361} (\bibinfo {year}
  {1985})}\BibitemShut {NoStop}%
%%CITATION = PHLTA,B164,361;%%
\bibitem [{\citenamefont {Krehl}\ \emph {et~al.}(2000)\citenamefont {Krehl},
  \citenamefont {Hanhart}, \citenamefont {Krewald},\ and\ \citenamefont
  {Speth}}]{Krehl:1999km}%
  \BibitemOpen
  \bibfield  {author} {\bibinfo {author} {\bibfnamefont {O.}~\bibnamefont
  {Krehl}}, \bibinfo {author} {\bibfnamefont {C.}~\bibnamefont {Hanhart}},
  \bibinfo {author} {\bibfnamefont {S.}~\bibnamefont {Krewald}}, \ and\
  \bibinfo {author} {\bibfnamefont {J.}~\bibnamefont {Speth}},\ }\href
  {\doibase 10.1103/PhysRevC.62.025207} {\bibfield  {journal} {\bibinfo
  {journal} {Phys. Rev.}\ }\textbf {\bibinfo {volume} {{C62}}},\ \bibinfo
  {pages} {025207} (\bibinfo {year} {2000})},\ \Eprint
  {http://arxiv.org/abs/nucl-th/9911080} {arXiv:nucl-th/9911080} \BibitemShut
  {NoStop}%
%%CITATION = NUCL-TH/9911080;%%
\bibitem [{\citenamefont {Allton}\ \emph {et~al.}(1993)\citenamefont {Allton}
  \emph {et~al.}}]{Allton:1993wc}%
  \BibitemOpen
  \bibfield  {author} {\bibinfo {author} {\bibfnamefont {C.~R.}\ \bibnamefont
  {Allton}} \emph {et~al.} (\bibinfo {collaboration} {UKQCD}),\ }\href@noop {}
  {\bibfield  {journal} {\bibinfo  {journal} {Phys. Rev.}\ }\textbf {\bibinfo
  {volume} {D47}},\ \bibinfo {pages} {5128} (\bibinfo {year} {1993})},\ \Eprint
  {http://arxiv.org/abs/hep-lat/9303009} {hep-lat/9303009} \BibitemShut
  {NoStop}%
%%CITATION = HEP-LAT/9303009;%%
\bibitem [{\citenamefont {Lee}\ and\ \citenamefont
  {Leinweber}(1999)}]{Lee:1998cx}%
  \BibitemOpen
  \bibfield  {author} {\bibinfo {author} {\bibfnamefont {F.~X.}\ \bibnamefont
  {Lee}}\ and\ \bibinfo {author} {\bibfnamefont {D.~B.}\ \bibnamefont
  {Leinweber}},\ }\href {\doibase 10.1016/S0920-5632(99)85041-5} {\bibfield
  {journal} {\bibinfo  {journal} {Nucl. Phys. Proc. Suppl.}\ }\textbf {\bibinfo
  {volume} {73}},\ \bibinfo {pages} {258} (\bibinfo {year} {1999})},\ \Eprint
  {http://arxiv.org/abs/hep-lat/9809095} {arXiv:hep-lat/9809095} \BibitemShut
  {NoStop}%
%%CITATION = HEP-LAT/9809095;%%
\bibitem [{\citenamefont {Gockeler}\ \emph {et~al.}(2002)\citenamefont
  {Gockeler} \emph {et~al.}}]{Gockeler:2001db}%
  \BibitemOpen
  \bibfield  {author} {\bibinfo {author} {\bibfnamefont {M.}~\bibnamefont
  {Gockeler}} \emph {et~al.} (\bibinfo {collaboration} {QCDSF}),\ }\href
  {\doibase 10.1016/S0370-2693(02)01492-2} {\bibfield  {journal} {\bibinfo
  {journal} {Phys. Lett.}\ }\textbf {\bibinfo {volume} {B532}},\ \bibinfo
  {pages} {63} (\bibinfo {year} {2002})},\ \Eprint
  {http://arxiv.org/abs/hep-lat/0106022} {arXiv:hep-lat/0106022} \BibitemShut
  {NoStop}%
%%CITATION = HEP-LAT/0106022;%%
\bibitem [{\citenamefont {Sasaki}\ \emph {et~al.}(2002)\citenamefont {Sasaki},
  \citenamefont {Blum},\ and\ \citenamefont {Ohta}}]{Sasaki:2001nf}%
  \BibitemOpen
  \bibfield  {author} {\bibinfo {author} {\bibfnamefont {S.}~\bibnamefont
  {Sasaki}}, \bibinfo {author} {\bibfnamefont {T.}~\bibnamefont {Blum}}, \ and\
  \bibinfo {author} {\bibfnamefont {S.}~\bibnamefont {Ohta}},\ }\href@noop {}
  {\bibfield  {journal} {\bibinfo  {journal} {Phys. Rev.}\ }\textbf {\bibinfo
  {volume} {D65}},\ \bibinfo {pages} {074503} (\bibinfo {year} {2002})},\
  \Eprint {http://arxiv.org/abs/hep-lat/0102010} {hep-lat/0102010} \BibitemShut
  {NoStop}%
%%CITATION = HEP-LAT/0102010;%%
\bibitem [{\citenamefont {Melnitchouk}\ \emph {et~al.}(2003)\citenamefont
  {Melnitchouk} \emph {et~al.}}]{Melnitchouk:2002eg}%
  \BibitemOpen
  \bibfield  {author} {\bibinfo {author} {\bibfnamefont {W.}~\bibnamefont
  {Melnitchouk}} \emph {et~al.},\ }\href@noop {} {\bibfield  {journal}
  {\bibinfo  {journal} {Phys. Rev.}\ }\textbf {\bibinfo {volume} {D67}},\
  \bibinfo {pages} {114506} (\bibinfo {year} {2003})},\ \Eprint
  {http://arxiv.org/abs/hep-lat/0202022} {hep-lat/0202022} \BibitemShut
  {NoStop}%
%%CITATION = HEP-LAT/0202022;%%
\bibitem [{\citenamefont {Edwards}\ \emph {et~al.}(2003)\citenamefont
  {Edwards}, \citenamefont {Heller},\ and\ \citenamefont
  {Richards}}]{Edwards:2003cd}%
  \BibitemOpen
  \bibfield  {author} {\bibinfo {author} {\bibfnamefont {R.~G.}\ \bibnamefont
  {Edwards}}, \bibinfo {author} {\bibfnamefont {U.~M.}\ \bibnamefont {Heller}},
  \ and\ \bibinfo {author} {\bibfnamefont {D.~G.}\ \bibnamefont {Richards}}
  (\bibinfo {collaboration} {LHP}),\ }\href {\doibase
  10.1016/S0920-5632(03)01525-1} {\bibfield  {journal} {\bibinfo  {journal}
  {Nucl. Phys. Proc. Suppl.}\ }\textbf {\bibinfo {volume} {119}},\ \bibinfo
  {pages} {305} (\bibinfo {year} {2003})},\ \Eprint
  {http://arxiv.org/abs/hep-lat/0303004} {arXiv:hep-lat/0303004} \BibitemShut
  {NoStop}%
%%CITATION = HEP-LAT/0303004;%%
\bibitem [{\citenamefont {Lee}\ \emph {et~al.}(2003)\citenamefont {Lee} \emph
  {et~al.}}]{Lee:2002gn}%
  \BibitemOpen
  \bibfield  {author} {\bibinfo {author} {\bibfnamefont {F.~X.}\ \bibnamefont
  {Lee}} \emph {et~al.},\ }\href@noop {} {\bibfield  {journal} {\bibinfo
  {journal} {Nucl. Phys. Proc. Suppl.}\ }\textbf {\bibinfo {volume} {119}},\
  \bibinfo {pages} {296} (\bibinfo {year} {2003})},\ \Eprint
  {http://arxiv.org/abs/hep-lat/0208070} {hep-lat/0208070} \BibitemShut
  {NoStop}%
%%CITATION = HEP-LAT/0208070;%%
\bibitem [{\citenamefont {Brommel}\ \emph {et~al.}(2004)\citenamefont {Brommel}
  \emph {et~al.}}]{Brommel:2003jm}%
  \BibitemOpen
  \bibfield  {author} {\bibinfo {author} {\bibfnamefont {D.}~\bibnamefont
  {Brommel}} \emph {et~al.} (\bibinfo {collaboration} {Bern-Graz-Regensburg}),\
  }\href@noop {} {\bibfield  {journal} {\bibinfo  {journal} {Phys. Rev.}\
  }\textbf {\bibinfo {volume} {D69}},\ \bibinfo {pages} {094513} (\bibinfo
  {year} {2004})},\ \Eprint {http://arxiv.org/abs/hep-ph/0307073}
  {hep-ph/0307073} \BibitemShut {NoStop}%
%%CITATION = HEP-PH/0307073;%%
\bibitem [{\citenamefont {Mathur}\ \emph {et~al.}(2005)\citenamefont {Mathur}
  \emph {et~al.}}]{Mathur:2003zf}%
  \BibitemOpen
  \bibfield  {author} {\bibinfo {author} {\bibfnamefont {N.}~\bibnamefont
  {Mathur}} \emph {et~al.},\ }\href@noop {} {\bibfield  {journal} {\bibinfo
  {journal} {Phys. Lett.}\ }\textbf {\bibinfo {volume} {B605}},\ \bibinfo
  {pages} {137} (\bibinfo {year} {2005})},\ \Eprint
  {http://arxiv.org/abs/hep-ph/0306199} {hep-ph/0306199} \BibitemShut {NoStop}%
%%CITATION = HEP-PH/0306199;%%
\bibitem [{\citenamefont {Sasaki}(2003)}]{Sasaki:2003xc}%
  \BibitemOpen
  \bibfield  {author} {\bibinfo {author} {\bibfnamefont {S.}~\bibnamefont
  {Sasaki}},\ }\href@noop {} {\bibfield  {journal} {\bibinfo  {journal} {Prog.
  Theor. Phys. Suppl.}\ }\textbf {\bibinfo {volume} {151}},\ \bibinfo {pages}
  {143} (\bibinfo {year} {2003})},\ \Eprint
  {http://arxiv.org/abs/nucl-th/0305014} {arXiv:nucl-th/0305014} \BibitemShut
  {NoStop}%
%%CITATION = NUCL-TH/0305014;%%
\bibitem [{\citenamefont {Burch}\ \emph {et~al.}(2004)\citenamefont {Burch}
  \emph {et~al.}}]{Burch:2004he}%
  \BibitemOpen
  \bibfield  {author} {\bibinfo {author} {\bibfnamefont {T.}~\bibnamefont
  {Burch}} \emph {et~al.} (\bibinfo {collaboration} {Bern-Graz-Regensburg}),\
  }\href@noop {} {\bibfield  {journal} {\bibinfo  {journal} {Phys. Rev.}\
  }\textbf {\bibinfo {volume} {D70}},\ \bibinfo {pages} {054502} (\bibinfo
  {year} {2004})},\ \Eprint {http://arxiv.org/abs/hep-lat/0405006}
  {hep-lat/0405006} \BibitemShut {NoStop}%
%%CITATION = HEP-LAT/0405006;%%
\bibitem [{\citenamefont {Basak}\ \emph {et~al.}(2007)\citenamefont {Basak}
  \emph {et~al.}}]{Basak:2007kj}%
  \BibitemOpen
  \bibfield  {author} {\bibinfo {author} {\bibfnamefont {S.}~\bibnamefont
  {Basak}} \emph {et~al.},\ }\href@noop {} {\bibfield  {journal} {\bibinfo
  {journal} {Phys. Rev.}\ }\textbf {\bibinfo {volume} {D76}},\ \bibinfo {pages}
  {074504} (\bibinfo {year} {2007})},\ \Eprint
  {http://arxiv.org/abs/arXiv:0709.0008 [hep-lat]} {arXiv:0709.0008 [hep-lat]}
  \BibitemShut {NoStop}%
%%CITATION = ARXIV:0709.0008;%%
\bibitem [{\citenamefont {Mahbub}\ \emph
  {et~al.}(2009{\natexlab{a}})\citenamefont {Mahbub} \emph
  {et~al.}}]{Mahbub:2009nr}%
  \BibitemOpen
  \bibfield  {author} {\bibinfo {author} {\bibfnamefont {M.~S.}\ \bibnamefont
  {Mahbub}} \emph {et~al.},\ }\href {\doibase 10.1103/PhysRevD.80.054507}
  {\bibfield  {journal} {\bibinfo  {journal} {Phys. Rev.}\ }\textbf {\bibinfo
  {volume} {{D80}}},\ \bibinfo {pages} {054507} (\bibinfo {year}
  {2009}{\natexlab{a}})},\ \Eprint {http://arxiv.org/abs/0905.3616}
  {arXiv:0905.3616 [hep-lat]} \BibitemShut {NoStop}%
%%CITATION = 0905.3616;%%
\bibitem [{\citenamefont {Mahbub}\ \emph
  {et~al.}(2009{\natexlab{b}})\citenamefont {Mahbub} \emph
  {et~al.}}]{Mahbub:2009aa}%
  \BibitemOpen
  \bibfield  {author} {\bibinfo {author} {\bibfnamefont {M.~S.}\ \bibnamefont
  {Mahbub}} \emph {et~al.},\ }\href {\doibase 10.1016/j.physletb.2009.07.063}
  {\bibfield  {journal} {\bibinfo  {journal} {Phys. Lett.}\ }\textbf {\bibinfo
  {volume} {B679}},\ \bibinfo {pages} {418} (\bibinfo {year}
  {2009}{\natexlab{b}})},\ \Eprint {http://arxiv.org/abs/0906.5433}
  {arXiv:0906.5433 [hep-lat]} \BibitemShut {NoStop}%
%%CITATION = 0906.5433;%%
\bibitem [{\citenamefont {Fleming}\ \emph {et~al.}(2009)\citenamefont
  {Fleming}, \citenamefont {Cohen}, \citenamefont {Lin},\ and\ \citenamefont
  {Pereyra}}]{Fleming:2009wb}%
  \BibitemOpen
  \bibfield  {author} {\bibinfo {author} {\bibfnamefont {G.~T.}\ \bibnamefont
  {Fleming}}, \bibinfo {author} {\bibfnamefont {S.~D.}\ \bibnamefont {Cohen}},
  \bibinfo {author} {\bibfnamefont {H.-W.}\ \bibnamefont {Lin}}, \ and\
  \bibinfo {author} {\bibfnamefont {V.}~\bibnamefont {Pereyra}},\ }\href
  {\doibase 10.1103/PhysRevD.80.074506} {\bibfield  {journal} {\bibinfo
  {journal} {Phys. Rev.}\ }\textbf {\bibinfo {volume} {D80}},\ \bibinfo {pages}
  {074506} (\bibinfo {year} {2009})},\ \Eprint {http://arxiv.org/abs/0903.2314}
  {arXiv:0903.2314 [hep-lat]} \BibitemShut {NoStop}%
%%CITATION = 0903.2314;%%
\bibitem [{\citenamefont {Mahbub}\ \emph
  {et~al.}(2010{\natexlab{a}})\citenamefont {Mahbub}, \citenamefont {Cais},
  \citenamefont {Kamleh}, \citenamefont {Leinweber},\ and\ \citenamefont
  {Williams}}]{Mahbub:2010jz}%
  \BibitemOpen
  \bibfield  {author} {\bibinfo {author} {\bibfnamefont {M.~S.}\ \bibnamefont
  {Mahbub}}, \bibinfo {author} {\bibfnamefont {A.~O.}\ \bibnamefont {Cais}},
  \bibinfo {author} {\bibfnamefont {W.}~\bibnamefont {Kamleh}}, \bibinfo
  {author} {\bibfnamefont {D.~B.}\ \bibnamefont {Leinweber}}, \ and\ \bibinfo
  {author} {\bibfnamefont {A.~G.}\ \bibnamefont {Williams}},\ }\href@noop {}
  {\bibfield  {journal} {\bibinfo  {journal} {Phys. Rev.}\ }\textbf {\bibinfo
  {volume} {D82}},\ \bibinfo {pages} {094504} (\bibinfo {year}
  {2010}{\natexlab{a}})},\ \Eprint {http://arxiv.org/abs/1004.5455}
  {arXiv:1004.5455 [hep-lat]} \BibitemShut {NoStop}%
%%CITATION = 1004.5455;%%
\bibitem [{\citenamefont {Mahbub}\ \emph
  {et~al.}(2010{\natexlab{b}})\citenamefont {Mahbub}, \citenamefont {Kamleh},
  \citenamefont {Leinweber}, \citenamefont {Cais},\ and\ \citenamefont
  {Williams}}]{Mahbub:2010me}%
  \BibitemOpen
  \bibfield  {author} {\bibinfo {author} {\bibfnamefont {M.~S.}\ \bibnamefont
  {Mahbub}}, \bibinfo {author} {\bibfnamefont {W.}~\bibnamefont {Kamleh}},
  \bibinfo {author} {\bibfnamefont {D.~B.}\ \bibnamefont {Leinweber}}, \bibinfo
  {author} {\bibfnamefont {A.~O.}\ \bibnamefont {Cais}}, \ and\ \bibinfo
  {author} {\bibfnamefont {A.~G.}\ \bibnamefont {Williams}},\ }\href {\doibase
  10.1016/j.physletb.2010.08.049} {\bibfield  {journal} {\bibinfo  {journal}
  {Phys. Lett.}\ }\textbf {\bibinfo {volume} {B693}},\ \bibinfo {pages} {351}
  (\bibinfo {year} {2010}{\natexlab{b}})},\ \Eprint
  {http://arxiv.org/abs/1007.4871} {arXiv:1007.4871 [hep-lat]} \BibitemShut
  {NoStop}%
%%CITATION = 1007.4871;%%
\bibitem [{\citenamefont {Bulava}\ \emph {et~al.}(2009)\citenamefont {Bulava}
  \emph {et~al.}}]{Bulava:2009jb}%
  \BibitemOpen
  \bibfield  {author} {\bibinfo {author} {\bibfnamefont {J.~M.}\ \bibnamefont
  {Bulava}} \emph {et~al.},\ }\href {\doibase 10.1103/PhysRevD.79.034505}
  {\bibfield  {journal} {\bibinfo  {journal} {Phys. Rev.}\ }\textbf {\bibinfo
  {volume} {D79}},\ \bibinfo {pages} {034505} (\bibinfo {year} {2009})},\
  \Eprint {http://arxiv.org/abs/0901.0027} {arXiv:0901.0027 [hep-lat]}
  \BibitemShut {NoStop}%
%%CITATION = 0901.0027;%%
\bibitem [{\citenamefont {Engel}\ \emph {et~al.}(2010)\citenamefont {Engel},
  \citenamefont {Lang}, \citenamefont {Limmer}, \citenamefont {Mohler},\ and\
  \citenamefont {Schafer}}]{Engel:2010my}%
  \BibitemOpen
  \bibfield  {author} {\bibinfo {author} {\bibfnamefont {G.~P.}\ \bibnamefont
  {Engel}}, \bibinfo {author} {\bibfnamefont {C.~B.}\ \bibnamefont {Lang}},
  \bibinfo {author} {\bibfnamefont {M.}~\bibnamefont {Limmer}}, \bibinfo
  {author} {\bibfnamefont {D.}~\bibnamefont {Mohler}}, \ and\ \bibinfo {author}
  {\bibfnamefont {A.}~\bibnamefont {Schafer}} (\bibinfo {collaboration} {BGR
  [Bern-Graz-Regensburg]}),\ }\href {\doibase 10.1103/PhysRevD.82.034505}
  {\bibfield  {journal} {\bibinfo  {journal} {Phys. Rev.}\ }\textbf {\bibinfo
  {volume} {D82}},\ \bibinfo {pages} {034505} (\bibinfo {year} {2010})},\
  \Eprint {http://arxiv.org/abs/1005.1748} {arXiv:1005.1748 [hep-lat]}
  \BibitemShut {NoStop}%
%%CITATION = 1005.1748;%%
\bibitem [{\citenamefont {Edwards}\ \emph {et~al.}(2011)\citenamefont
  {Edwards}, \citenamefont {Dudek}, \citenamefont {Richards},\ and\
  \citenamefont {Wallace}}]{Edwards:2011jj}%
  \BibitemOpen
  \bibfield  {author} {\bibinfo {author} {\bibfnamefont {R.~G.}\ \bibnamefont
  {Edwards}}, \bibinfo {author} {\bibfnamefont {J.~J.}\ \bibnamefont {Dudek}},
  \bibinfo {author} {\bibfnamefont {D.~G.}\ \bibnamefont {Richards}}, \ and\
  \bibinfo {author} {\bibfnamefont {S.~J.}\ \bibnamefont {Wallace}},\ }\href
  {\doibase 10.1103/PhysRevD.84.074508} {\bibfield  {journal} {\bibinfo
  {journal} {Phys. Rev.}\ }\textbf {\bibinfo {volume} {D84}},\ \bibinfo {pages}
  {074508} (\bibinfo {year} {2011})},\ \Eprint {http://arxiv.org/abs/1104.5152}
  {arXiv:1104.5152 [hep-ph]} \BibitemShut {NoStop}%
%%CITATION = 1104.5152;%%
\bibitem [{\citenamefont {Mahbub}\ \emph {et~al.}(2012)\citenamefont {Mahbub},
  \citenamefont {Kamleh}, \citenamefont {Leinweber}, \citenamefont {Moran},\
  and\ \citenamefont {Williams}}]{Mahbub:2010rm}%
  \BibitemOpen
  \bibfield  {author} {\bibinfo {author} {\bibfnamefont {M.~S.}\ \bibnamefont
  {Mahbub}}, \bibinfo {author} {\bibfnamefont {W.}~\bibnamefont {Kamleh}},
  \bibinfo {author} {\bibfnamefont {D.~B.}\ \bibnamefont {Leinweber}}, \bibinfo
  {author} {\bibfnamefont {P.~J.}\ \bibnamefont {Moran}}, \ and\ \bibinfo
  {author} {\bibfnamefont {A.~G.}\ \bibnamefont {Williams}} (\bibinfo
  {collaboration} {CSSM Lattice}),\ }\href@noop {} {\bibfield  {journal}
  {\bibinfo  {journal} {Phys. Lett.}\ }\textbf {\bibinfo {volume} {B707}},\
  \bibinfo {pages} {389} (\bibinfo {year} {2012})},\ \Eprint
  {http://arxiv.org/abs/1011.5724} {arXiv:1011.5724 [hep-lat]} \BibitemShut
  {NoStop}%
%%CITATION = 1011.5724;%%
\bibitem [{\citenamefont {Lin}\ and\ \citenamefont {Cohen}(2011)}]{Lin:2011da}%
  \BibitemOpen
  \bibfield  {author} {\bibinfo {author} {\bibfnamefont {H.-W.}\ \bibnamefont
  {Lin}}\ and\ \bibinfo {author} {\bibfnamefont {S.~D.}\ \bibnamefont
  {Cohen}},\ }\href@noop {} {\  (\bibinfo {year} {2011})},\ \Eprint
  {http://arxiv.org/abs/1108.2528} {arXiv:1108.2528 [hep-lat]} \BibitemShut
  {NoStop}%
%%CITATION = 1108.2528;%%
\bibitem [{\citenamefont {Michael}(1985)}]{Michael:1985ne}%
  \BibitemOpen
  \bibfield  {author} {\bibinfo {author} {\bibfnamefont {C.}~\bibnamefont
  {Michael}},\ }\href@noop {} {\bibfield  {journal} {\bibinfo  {journal} {Nucl.
  Phys.}\ }\textbf {\bibinfo {volume} {B259}},\ \bibinfo {pages} {58} (\bibinfo
  {year} {1985})}\BibitemShut {NoStop}%
%%CITATION = NUPHA,B259,58;%%
\bibitem [{\citenamefont {Luscher}\ and\ \citenamefont
  {Wolff}(1990)}]{Luscher:1990ck}%
  \BibitemOpen
  \bibfield  {author} {\bibinfo {author} {\bibfnamefont {M.}~\bibnamefont
  {Luscher}}\ and\ \bibinfo {author} {\bibfnamefont {U.}~\bibnamefont
  {Wolff}},\ }\href@noop {} {\bibfield  {journal} {\bibinfo  {journal} {Nucl.
  Phys.}\ }\textbf {\bibinfo {volume} {B339}},\ \bibinfo {pages} {222}
  (\bibinfo {year} {1990})}\BibitemShut {NoStop}%
%%CITATION = NUPHA,B339,222;%%
\bibitem [{\citenamefont {Lin}\ \emph {et~al.}(2009)\citenamefont {Lin} \emph
  {et~al.}}]{Lin:2008pr}%
  \BibitemOpen
  \bibfield  {author} {\bibinfo {author} {\bibfnamefont {H.-W.}\ \bibnamefont
  {Lin}} \emph {et~al.} (\bibinfo {collaboration} {Hadron Spectrum}),\ }\href
  {\doibase 10.1103/PhysRevD.79.034502} {\bibfield  {journal} {\bibinfo
  {journal} {Phys. Rev.}\ }\textbf {\bibinfo {volume} {D79}},\ \bibinfo {pages}
  {034502} (\bibinfo {year} {2009})},\ \Eprint {http://arxiv.org/abs/0810.3588}
  {arXiv:0810.3588 [hep-lat]} \BibitemShut {NoStop}%
%%CITATION = 0810.3588;%%
\bibitem [{\citenamefont {Bulava}\ \emph {et~al.}(2010)\citenamefont {Bulava}
  \emph {et~al.}}]{Bulava:2010yg}%
  \BibitemOpen
  \bibfield  {author} {\bibinfo {author} {\bibfnamefont {J.}~\bibnamefont
  {Bulava}} \emph {et~al.},\ }\href {\doibase 10.1103/PhysRevD.82.014507}
  {\bibfield  {journal} {\bibinfo  {journal} {Phys. Rev.}\ }\textbf {\bibinfo
  {volume} {D82}},\ \bibinfo {pages} {014507} (\bibinfo {year} {2010})},\
  \Eprint {http://arxiv.org/abs/1004.5072} {arXiv:1004.5072 [hep-lat]}
  \BibitemShut {NoStop}%
%%CITATION = 1004.5072;%%
\bibitem [{\citenamefont {Morningstar}\ \emph {et~al.}(2011)\citenamefont
  {Morningstar} \emph {et~al.}}]{Morningstar:2011ka}%
  \BibitemOpen
  \bibfield  {author} {\bibinfo {author} {\bibfnamefont {C.}~\bibnamefont
  {Morningstar}} \emph {et~al.},\ }\href {\doibase 10.1103/PhysRevD.83.114505}
  {\bibfield  {journal} {\bibinfo  {journal} {Phys. Rev.}\ }\textbf {\bibinfo
  {volume} {D83}},\ \bibinfo {pages} {114505} (\bibinfo {year} {2011})},\
  \Eprint {http://arxiv.org/abs/1104.3870} {arXiv:1104.3870 [hep-lat]}
  \BibitemShut {NoStop}%
%%CITATION = 1104.3870;%%
\bibitem [{\citenamefont {Bulava}(2011)}]{Bulava:2011uk}%
  \BibitemOpen
  \bibfield  {author} {\bibinfo {author} {\bibfnamefont {J.}~\bibnamefont
  {Bulava}},\ }\href@noop {} {\  (\bibinfo {year} {2011})},\ \Eprint
  {http://arxiv.org/abs/1112.0212} {arXiv:1112.0212 [hep-lat]} \BibitemShut
  {NoStop}%
%%CITATION = 1112.0212;%%
\bibitem [{\citenamefont {Menadue}\ \emph {et~al.}(2011)\citenamefont
  {Menadue}, \citenamefont {Kamleh}, \citenamefont {Leinweber},\ and\
  \citenamefont {Mahbub}}]{Menadue:2011pd}%
  \BibitemOpen
  \bibfield  {author} {\bibinfo {author} {\bibfnamefont {B.~J.}\ \bibnamefont
  {Menadue}}, \bibinfo {author} {\bibfnamefont {W.}~\bibnamefont {Kamleh}},
  \bibinfo {author} {\bibfnamefont {D.~B.}\ \bibnamefont {Leinweber}}, \ and\
  \bibinfo {author} {\bibfnamefont {M.~S.}\ \bibnamefont {Mahbub}},\
  }\href@noop {} {\  (\bibinfo {year} {2011})},\ \Eprint
  {http://arxiv.org/abs/1109.6716} {arXiv:1109.6716 [hep-lat]} \BibitemShut
  {NoStop}%
%%CITATION = 1109.6716;%%
\bibitem [{\citenamefont {Edwards}\ \emph {et~al.}(2012)\citenamefont
  {Edwards}, \citenamefont {Mathur}, \citenamefont {Richards},\ and\
  \citenamefont {Wallace}}]{Edwards:2012fx}%
  \BibitemOpen
  \bibfield  {author} {\bibinfo {author} {\bibfnamefont {R.~G.}\ \bibnamefont
  {Edwards}}, \bibinfo {author} {\bibfnamefont {N.}~\bibnamefont {Mathur}},
  \bibinfo {author} {\bibfnamefont {D.~G.}\ \bibnamefont {Richards}}, \ and\
  \bibinfo {author} {\bibfnamefont {S.~J.}\ \bibnamefont {Wallace}},\
  }\href@noop {} {\  (\bibinfo {year} {2012})},\ \Eprint
  {http://arxiv.org/abs/1212.5236} {arXiv:1212.5236 [hep-ph]} \BibitemShut
  {NoStop}%
%%CITATION = ARXIV:1212.5236;%%
\bibitem [{\citenamefont {Engel}\ \emph {et~al.}(2013)\citenamefont {Engel},
  \citenamefont {Lang}, \citenamefont {Mohler},\ and\ \citenamefont
  {Schaefer}}]{Engel:2013ig}%
  \BibitemOpen
  \bibfield  {author} {\bibinfo {author} {\bibfnamefont {G.~P.}\ \bibnamefont
  {Engel}}, \bibinfo {author} {\bibfnamefont {C.}~\bibnamefont {Lang}},
  \bibinfo {author} {\bibfnamefont {D.}~\bibnamefont {Mohler}}, \ and\ \bibinfo
  {author} {\bibfnamefont {A.}~\bibnamefont {Schaefer}},\ }\href@noop {} {\
  (\bibinfo {year} {2013})},\ \Eprint {http://arxiv.org/abs/1301.4318}
  {arXiv:1301.4318 [hep-lat]} \BibitemShut {NoStop}%
%%CITATION = ARXIV:1301.4318;%%
\bibitem [{\citenamefont {Aoki}\ \emph {et~al.}(2009)\citenamefont {Aoki} \emph
  {et~al.}}]{Aoki:2008sm}%
  \BibitemOpen
  \bibfield  {author} {\bibinfo {author} {\bibfnamefont {S.}~\bibnamefont
  {Aoki}} \emph {et~al.} (\bibinfo {collaboration} {PACS-CS}),\ }\href@noop {}
  {\bibfield  {journal} {\bibinfo  {journal} {Phys. Rev. D}\ }\textbf {\bibinfo
  {volume} {79}},\ \bibinfo {pages} {034503} (\bibinfo {year} {2009})},\
  \Eprint {http://arxiv.org/abs/0807.1661} {arXiv:0807.1661 [hep-lat]}
  \BibitemShut {NoStop}%
%%CITATION = 0807.1661;%%
\bibitem [{\citenamefont {Mahbub}\ \emph {et~al.}(2013)\citenamefont {Mahbub},
  \citenamefont {Kamleh}, \citenamefont {Leinweber}, \citenamefont {Moran},\
  and\ \citenamefont {Williams}}]{Mahbub:2012ri}%
  \BibitemOpen
  \bibfield  {author} {\bibinfo {author} {\bibfnamefont {M.~S.}\ \bibnamefont
  {Mahbub}}, \bibinfo {author} {\bibfnamefont {W.}~\bibnamefont {Kamleh}},
  \bibinfo {author} {\bibfnamefont {D.~B.}\ \bibnamefont {Leinweber}}, \bibinfo
  {author} {\bibfnamefont {P.~J.}\ \bibnamefont {Moran}}, \ and\ \bibinfo
  {author} {\bibfnamefont {A.~G.}\ \bibnamefont {Williams}} (\bibinfo
  {collaboration} {CSSM Lattice Collaboration}),\ }\href {\doibase
  10.1103/PhysRevD.87.011501} {\bibfield  {journal} {\bibinfo  {journal}
  {Phys.Rev.}\ }\textbf {\bibinfo {volume} {D87}},\ \bibinfo {pages} {011501}
  (\bibinfo {year} {2013})},\ \Eprint {http://arxiv.org/abs/1209.0240}
  {arXiv:1209.0240 [hep-lat]} \BibitemShut {NoStop}%
%%CITATION = ARXIV:1209.0240;%%
\bibitem [{\citenamefont {Lang}\ and\ \citenamefont
  {Verduci}(2012)}]{Lang:2012db}%
  \BibitemOpen
  \bibfield  {author} {\bibinfo {author} {\bibfnamefont {C.}~\bibnamefont
  {Lang}}\ and\ \bibinfo {author} {\bibfnamefont {V.}~\bibnamefont {Verduci}},\
  }\href@noop {} {\  (\bibinfo {year} {2012})},\ \Eprint
  {http://arxiv.org/abs/1212.5055} {arXiv:1212.5055 [hep-lat]} \BibitemShut
  {NoStop}%
%%CITATION = ARXIV:1212.5055;%%
\bibitem [{\citenamefont {Morningstar}\ \emph {et~al.}(2013)\citenamefont
  {Morningstar}, \citenamefont {Bulava}, \citenamefont {Fahy}, \citenamefont
  {Foley}, \citenamefont {Jhang} \emph {et~al.}}]{Morningstar:2013bda}%
  \BibitemOpen
  \bibfield  {author} {\bibinfo {author} {\bibfnamefont {C.}~\bibnamefont
  {Morningstar}}, \bibinfo {author} {\bibfnamefont {J.}~\bibnamefont {Bulava}},
  \bibinfo {author} {\bibfnamefont {B.}~\bibnamefont {Fahy}}, \bibinfo {author}
  {\bibfnamefont {J.}~\bibnamefont {Foley}}, \bibinfo {author} {\bibfnamefont
  {Y.}~\bibnamefont {Jhang}},  \emph {et~al.},\ }\href@noop {} {\  (\bibinfo
  {year} {2013})},\ \Eprint {http://arxiv.org/abs/1303.6816} {arXiv:1303.6816
  [hep-lat]} \BibitemShut {NoStop}%
%%CITATION = ARXIV:1303.6816;%%
\bibitem [{\citenamefont {Gusken}(1990)}]{Gusken:1989qx}%
  \BibitemOpen
  \bibfield  {author} {\bibinfo {author} {\bibfnamefont {S.}~\bibnamefont
  {Gusken}},\ }\href@noop {} {\bibfield  {journal} {\bibinfo  {journal} {Nucl.
  Phys. Proc. Suppl.}\ }\textbf {\bibinfo {volume} {17}},\ \bibinfo {pages}
  {361} (\bibinfo {year} {1990})}\BibitemShut {NoStop}%
%%CITATION = NUPHZ,17,361;%%
\bibitem [{\citenamefont {Blossier}\ \emph {et~al.}(2009)\citenamefont
  {Blossier}, \citenamefont {Della~Morte}, \citenamefont {von Hippel},
  \citenamefont {Mendes},\ and\ \citenamefont {Sommer}}]{Blossier:2009kd}%
  \BibitemOpen
  \bibfield  {author} {\bibinfo {author} {\bibfnamefont {B.}~\bibnamefont
  {Blossier}}, \bibinfo {author} {\bibfnamefont {M.}~\bibnamefont
  {Della~Morte}}, \bibinfo {author} {\bibfnamefont {G.}~\bibnamefont {von
  Hippel}}, \bibinfo {author} {\bibfnamefont {T.}~\bibnamefont {Mendes}}, \
  and\ \bibinfo {author} {\bibfnamefont {R.}~\bibnamefont {Sommer}},\ }\href
  {\doibase 10.1088/1126-6708/2009/04/094} {\bibfield  {journal} {\bibinfo
  {journal} {JHEP}\ }\textbf {\bibinfo {volume} {04}},\ \bibinfo {pages} {094}
  (\bibinfo {year} {2009})},\ \Eprint {http://arxiv.org/abs/0902.1265}
  {arXiv:0902.1265 [hep-lat]} \BibitemShut {NoStop}%
%%CITATION = 0902.1265;%%
\bibitem [{\citenamefont {Beckett}\ \emph {et~al.}(2011)\citenamefont {Beckett}
  \emph {et~al.}}]{Beckett:2009cb}%
  \BibitemOpen
  \bibfield  {author} {\bibinfo {author} {\bibfnamefont {M.~G.}\ \bibnamefont
  {Beckett}} \emph {et~al.},\ }\href {\doibase 10.1016/j.cpc.2011.01.027}
  {\bibfield  {journal} {\bibinfo  {journal} {Comput. Phys. Commun.}\ }\textbf
  {\bibinfo {volume} {182}},\ \bibinfo {pages} {1208} (\bibinfo {year}
  {2011})},\ \Eprint {http://arxiv.org/abs/0910.1692} {arXiv:0910.1692
  [hep-lat]} \BibitemShut {NoStop}%
%%CITATION = 0910.1692;%%
\bibitem [{\citenamefont {Iwasaki}(1983)}]{Iwasaki:1983ck}%
  \BibitemOpen
  \bibfield  {author} {\bibinfo {author} {\bibfnamefont {Y.}~\bibnamefont
  {Iwasaki}},\ }\href@noop {} {\  (\bibinfo {year} {1983})},\ \bibinfo {note}
  {uTHEP-118}\BibitemShut {NoStop}%
\bibitem [{\citenamefont {Leinweber}\ \emph {et~al.}(1991)\citenamefont
  {Leinweber}, \citenamefont {Woloshyn},\ and\ \citenamefont
  {Draper}}]{Leinweber:1990dv}%
  \BibitemOpen
  \bibfield  {author} {\bibinfo {author} {\bibfnamefont {D.~B.}\ \bibnamefont
  {Leinweber}}, \bibinfo {author} {\bibfnamefont {R.~M.}\ \bibnamefont
  {Woloshyn}}, \ and\ \bibinfo {author} {\bibfnamefont {T.}~\bibnamefont
  {Draper}},\ }\href@noop {} {\bibfield  {journal} {\bibinfo  {journal} {Phys.
  Rev.}\ }\textbf {\bibinfo {volume} {D43}},\ \bibinfo {pages} {1659} (\bibinfo
  {year} {1991})}\BibitemShut {NoStop}%
%%CITATION = PHRVA,D43,1659;%%
\bibitem [{\citenamefont {Leinweber}(1995)}]{Leinweber:1994nm}%
  \BibitemOpen
  \bibfield  {author} {\bibinfo {author} {\bibfnamefont {D.~B.}\ \bibnamefont
  {Leinweber}},\ }\href@noop {} {\bibfield  {journal} {\bibinfo  {journal}
  {Phys. Rev.}\ }\textbf {\bibinfo {volume} {D51}},\ \bibinfo {pages} {6383}
  (\bibinfo {year} {1995})},\ \Eprint {http://arxiv.org/abs/nucl-th/9406001}
  {nucl-th/9406001} \BibitemShut {NoStop}%
%%CITATION = NUCL-TH/9406001;%%
\bibitem [{\citenamefont {Zanotti}\ \emph {et~al.}(2003)\citenamefont {Zanotti}
  \emph {et~al.}}]{Zanotti:2003fx}%
  \BibitemOpen
  \bibfield  {author} {\bibinfo {author} {\bibfnamefont {J.~M.}\ \bibnamefont
  {Zanotti}} \emph {et~al.} (\bibinfo {collaboration} {CSSM Lattice}),\
  }\href@noop {} {\bibfield  {journal} {\bibinfo  {journal} {Phys. Rev.}\
  }\textbf {\bibinfo {volume} {D68}},\ \bibinfo {pages} {054506} (\bibinfo
  {year} {2003})},\ \Eprint {http://arxiv.org/abs/hep-lat/0304001}
  {hep-lat/0304001} \BibitemShut {NoStop}%
%%CITATION = HEP-LAT/0304001;%%
\bibitem [{\citenamefont {Young}\ \emph {et~al.}(2002)\citenamefont {Young},
  \citenamefont {Leinweber}, \citenamefont {Thomas},\ and\ \citenamefont
  {Wright}}]{Young:2002cj}%
  \BibitemOpen
  \bibfield  {author} {\bibinfo {author} {\bibfnamefont {R.~D.}\ \bibnamefont
  {Young}}, \bibinfo {author} {\bibfnamefont {D.~B.}\ \bibnamefont
  {Leinweber}}, \bibinfo {author} {\bibfnamefont {A.~W.}\ \bibnamefont
  {Thomas}}, \ and\ \bibinfo {author} {\bibfnamefont {S.~V.}\ \bibnamefont
  {Wright}},\ }\href {\doibase 10.1103/PhysRevD.66.094507} {\bibfield
  {journal} {\bibinfo  {journal} {Phys. Rev.}\ }\textbf {\bibinfo {volume}
  {D66}},\ \bibinfo {pages} {094507} (\bibinfo {year} {2002})},\ \Eprint
  {http://arxiv.org/abs/hep-lat/0205017} {arXiv:hep-lat/0205017} \BibitemShut
  {NoStop}%
%%CITATION = HEP-LAT/0205017;%%
\end{thebibliography}
\end{document}